\documentclass[fleqn,usenatbib]{mnras}

\usepackage{ae,aecompl}
\usepackage{xcolor}

\usepackage{graphicx}	% Including figure files
\usepackage{amsmath}	% Advanced maths commands
\usepackage{amssymb}	% Extra maths symbols
\usepackage{subfig}
\usepackage{mathtools}

\newcommand{\code}[1]{\texttt{#1}}
\newcommand*\dd{\mathop{}\!\mathrm{d}}

\begin{document}

\title[Radio Galaxy Evolution and Jet Magnetic Topology]{A Numerical Study of the Impact of Jet Magnetic Topology on Radio Galaxy Evolution}

\author[Yi-Hao Chen et al.]{
	Yi-Hao Chen,$^{1}$ and Sebastian Heinz$^{1}$\thanks{E-mail: sheinz@wisc.edu}, and Eric Hooper$^{1}$
\\
% List of institutions
$^{1}$Department of Astronomy, University of Wisconsin-Madison, 475 N. Charter Street, Madison, WI 53706, USA
}

% The list of authors, and the short list which is used in the headers.
% If you need two or more lines of authors, add an extra line using \newauthor

\date{Accepted XXX. Received YYY; in original form ZZZ}

% Enter the current year, for the copyright statements etc.
\pubyear{2022}

% Don't change these lines

\maketitle

\begin{abstract}
  The propagation of active galactic nucleus jets depends both on the environment into which they propagate and on their internal structure. To test the impact that different magnetic topologies have on the observable properties of radio galaxies on kpc scales, we conducted a series of magneto-hydrodynamic simulations of jets injected with different magnetic field configurations propagating into a gaseous atmosphere modeled on the Perseus cluster. The simulations show that the structure of the field affects the collimation and propagation of the jets on cluster scales and thus the morphology of the radio lobes inflated by the jets, due both to magnetic collimation and the development of dynamical instabilities in jets with different magnetic topologies.  In all cases, the simulations show a distinct reversal of the sychrotron spectral age gradient in the radio lobes about a dynamical time after the jets turn off due to large scale circulation inside the radio lobe, driven primarily by buoyancy, which could provide a way to constrain the age of radio sources in cluster environments without the need for detailed spectral modeling and thus constrain the radio mode feedback efficiency. We suggest a robust diagnostic to search for such age gradients in multi-frequency radio data.
\end{abstract}

\begin{keywords}
galaxies: active --- galaxies: clusters: intracluster medium --- galaxies: jets --- magnetohydrodynamics (MHD) --- methods: numerical
\end{keywords}

% Physics of Jets: launching, propagation, interaction
% Astrophysics: baryonic galaxy cluster evolution, ICM physics, cooling flow, Perseus cluster
% Galaxy AGN Feedback,
% Importance of magnetic fields,
% Computation: advance of computing power, advance of numerical MHD models and AMR integration.
% Importance of resolution, difficulties in memory and computing time, RMHD not available with AMR

% Observational tests

\section{Introduction}

The physics of relativistic jets launched by compact objects present an ongoing modeling challenge. It is generally accepted that jets are launched by extracting the rotational energy from the black hole \citep[][hereafter BZ]{blandford:77} or the accretion disk \citep{blandford:82} through the threading of magnetic fields into the event horizon or the disk. The jet power of the BZ mechanism depends on the magnetic flux and spin of the black hole, which are both still poorly constrained. Large-scale poloidal fields, which could be advected by the accretion flow from larger radii \citep{beckwith:09} or generated through a dynamo around the compact object \citep{tout:96,bugli:14}, are likely necessary to accelerate the jets. 

Twisting of the magnetic fields by the rotation of the black hole or the accretion disk naturally leads to the generation of toroidal fields \citep{tchekhovskoy:15}. It is thought that, in the vicinity of the black hole, the flow could be dominated by Poynting flux as in the original BZ mechanism and modeled by \cite{li:06} and \cite{guan:14}, in which the magnetic energy density is much larger than the kinetic and thermal energy. If the magnetic energy is transferred to the particles within the jets, the particles can be accelerated to highly relativistic velocity through a large range of scales \citep[e.g.][]{heinz:00,lyubarsky:09}. Collimation, which could by facilitated by the magnetohydrodynamic (MHD) structure of the jets \citep{nakamura:07,mckinney:09} and/or by external pressure such as disk winds \citep{porth:10}, helps the jets remain beam-like for up to hundreds of kpc.

The large scale nebulae inflated by propagating jets, generally referred to as radio lobes or radio galaxies, are commonly observed to have morphologies described within the framework proposed by \citet{fanaroff:74} \citep[see also][]{banfield:15}. Nearby, resolved sources are seen with rich structures inside their radio lobes \citep{leahy:96,miley:80}. The energetic feedback by these radio sources likely plays an important role in the evolution of galaxies and their environments \citep[see e.g.][for reviews]{fabian:12,mcnamara:12}.

The BZ mechanism requires strong poloidal fields threading the event horizon of the black hole to accelerate the jets. \cite{beckwith:08} tried different magnetic field structures, including poloidal, toroidal, and quadrupolar in global general relativistic magneto-hydrodynamic (GRMHD) simulations, and found that only predominantly poloidal magnetic fields can generate jets. In these jets, the pitch angle between the poloidal and toroidal magnetic fields is an essential factor during the propagation of the jets. \cite{bromberg:16} point out that the pitch angle depends on the external medium that confines the jets, and a denser medium will result in a smaller pitch angle, i.e. stronger toroidal fields. 

The rapid advance of GRMHD simulations has enabled some critical insights into jet acceleration and jet-disk coupling, revealing that jets are almost unavoidable consequences of accretion onto rapidly rotating black holes, and that the net magnetic flux through the disk plays a critical role in regulating accretion and jet formation. In particular, in the magnetically arrested disk (MAD) framework, the accretion flow can choke if the magnetic field exceeds a critical value, leading to conditions conducive to acceleration of highly relativistic jets. Still, the magnetic topology within the jet on scales of tens of kiloparsec remains largely unknown.

Observationally, \cite{gabuzda:15} found gradients of Faraday rotation measure across the jet structure in observations with the Very Long Baseline Array (VLBA). These gradients suggest the existence of toroidal components of the magnetic field at least in the central few 10-pc region. However, Faraday
rotation shows only magnetic fields along the line of sight and can not tell the fields perpendicular to it. \citet{pasetto:21} interpret the double-helical emission morphology of the M87 jet as indication of a helical field topology, though this is an indirect argument, rather than a direct measurement of the magnetic field orientation.

From simple flux freezing arguments, it might appear reasonable to assume that the poloidal component of the field decays away rapidly as the jet expands laterally once it has accelerated to its terminal velocity. However, jet stability arguments suggest \citep[e.g.][]{begelman:98,barniol-duran:17} that purely toroidal fields might not be consistent with the observed structure of jets. In addition, to infer the asymptotic field topology still requires knowledge of the acceleration scale, the injection field, and the overall lateral expansion factor, which in most jets is not directly observable and depends on the cross-field force balance of the jet and thus on the magnetic field itself.

Thus, it is fair to say that the detailed magnetic topology inside jets is not fully understood. The work presented here aims to investigate what imprint, if any, the magnetic topology of jets might have on the large scale propagation of jets and the overall dynamical properties of radio lobes they inflate.  Given the insurmountable scale discrepancy between jet acceleration and large scale propagation, we do not attempt to simulate the ab-initio acceleration and collimation process. Instead, we focus on the interaction of jets with ambient gas, assuming most of the magnetic energy has been converted to kinetic energy on scales smaller than the inlets of our injection nozzle, roughly about 100 pc away from the black hole. Although the energy flux is dominated by the kinetic energy in our simulations, magnetic fields are still dynamically important within the jet.

Numerical experiments investigating the propagation of AGN jets have proliferated in recent years, motivated by the realization that feedback by jets is crucial for the evolution of galaxies and their environments. For example, \cite{mendygral:11,mendygral:12} performed 3D non-relativistic MHD simulations of jets, similar to the ones presented here, however, with higher plasma $\beta$.  \cite{weinberger:17,bourne:21,smith:19,smith:21} present long-duration hydrodynamic simulations of cluster-scale feedback, focusing on the long-term co-evolution of jets and clusters. In a pioneering set of papers, \citet{horton:20,yates-jones:21} investigate the propagation of jets in clusters using relativistic hydrodynamic simulations, alleviating the shortcomings of sub-relativstic simulations typically employed in this context, including the work presented here (see \S\ref{sec:jet_injection} and \S\ref{sec:precession} for a discussion on the numerical necessity for this choice). In a complementary long-term study, \citet{hardcastle:13,hardcastle:14,english:16,english:19} present detailed magneto-hydrodynamic simulations of radio galaxies. All of these simulations approach the overall problem using different numerical methods and assumptions.

The work presented in this paper builds on and complements this already rich canon of published work, focusing on a specific question---the role played by magnetic topology in the inflation of radio lobes in galaxy clusters---rather than attempting to perform end-to-end simulations of cosmological structure formation with feedback. Specifically, we focus on jet propagation in the well-studied nearby Perseus cluster, with the expectation that findings are transferable to other cool core clusters at a minimum. 

The paper is organized as follows: In \S\ref{sec:numerical_techniques}, we describe the methodology and techniques used in the simulations. In \S\ref{sec:results}, we present the results of the simulations, including the dynamics and synchrotron spectral properties, \S\ref{sec:discussion} presents a discussion of the results, and \S\ref{sec:conclusions} summarizes our findings.

\section{Numerical Techniques}
\label{sec:numerical_techniques}

The scale-problem in jet modeling is well known: The physical launching radius of the jets is a few gravitational radii ($r_g = 2GM/c^2 = 10^{-4}$ pc for a $10^9 M_{\odot}$ black hole). The largest radio lobes can, however, extend to several Mpc \citep[e.g.,J1420-0545 and 3C 236,][]{machalski:08,tremblay:10}. It is currently impossible to simulate the acceleration of jets and the inflation of radio lobes together, which span over 10 orders of magnitude in spatial scales. Thus, studies of jets can be categorized into several regimes: (a) jet launching, (b) collimation, and (c) propagation, jet-ambient interaction. Due to the highly nonlinear physical processes and the complex and dynamic environments, applications of analytical solutions are usually limited. Thus, numerical simulations are necessary to study these systems. Here, we focus on simulations that model regimes (b) and (c). 

We use the grid-based adaptive mesh-refinement (AMR) magneto-hydrodynamic (MHD) code \code{FLASH}\footnote{\url{http://flash.uchicago.edu/}} \citep{fryxell:00,dubey:09} to perform the simulations. Full 3D simulations are conducted to model the non-linear nature of MHD. We include simulations of back-to-back jets injected into both hemispheres. This allows us to model jet-driven large scale flows across the mid-plane and to use a comparison of the two sides to estimate the uncertainty in derived quantities due to the stochasticity of non-linear fluid mechanics. The simulations under-gird and extend the work presented in \citet{chen:19} and \citet{heinrich:21} and follow the same overall setup.

We use the unsplit staggered mesh (USM) scheme implemented in FLASH by \cite{lee:13} for solving the MHD equations. The USM algorithm in \code{FLASH} is a finite-volume Godunov scheme employing the constrained transport method that preserves the divergence-free nature of magnetic field to machine precision. We use the hybrid Riemann solver, which combines the HLL and HLLD solvers for high accuracy and stability, 3rd- order reconstruction and the mc slope limiter. This solver is manifestly non-relativistic, the limitations and implications of which we discuss further below.

The initial conditions are described in detail in \S \ref{subsec:init_cond}. The simulation box is $1 \times 0.5 \times 0.5$ Mpc with $16 \times 8 \times 8$ base blocks and hydrostatic diode boundary conditions (\code{hydrostatic-f2+nvdiode}). Each AMR block contains $8^{3}$ cells. The minimum cell size is 30 pc under 12 levels of adaptive refinement. The jet nozzle diameter is resolved by 2 blocks, or 16 cells. The refinement criteria include the native second derivative based estimator and also a self-defined momentum-based condition that ensures that the jets are always covered at the highest level of refinement. We restrict the maximum refinement level for the ICM further away from the origin such that the number of blocks is always larger than 32 to the central plane (x-y plane) and larger than 16 to the the jet axis (z-axis).

The jets are active at a constant jet power of $L_{\rm jet} \approx 10^{45}\,{\rm ergs/s}$ for 10 Myr. The maximum refinement level (used to resolve the jets) is reduced after the jets are turned off, and the simulation is allowed to evolve passively for another 91 Myrs to study the long term evolution of radio lobes. While the choice of duty cycle is arbitrary, the size scale of the simulated radio lobes corresponds reasonably well with the observed cavity sizes in the centers of cool core clusters.

Most of the visualization and analysis work is performed with the \code{python}-based software package \code{yt}\footnote{\url{http://yt-project.org/}} \citep{turk:11}. The 3D rendering is carried out in \code{VisIt} \citep{childs:12}.

\begin{table}
	\caption{List of variables used in the simulation.}
	\label{tab:variables}
	\center
	\begin{tabular}{lccc}
		\hline
		\bf{Variable}		&	\bf{Symbol}		& \bf{Value}			& \bf{Unit}\\
		\hline
		jet power			&	$L$				&  $1\times10^{45}$ 	& erg/s\\
		\hline
		jet velocity		&	$v_j$			&  $3\times10^{9}$		& cm/s\\
		\hline
		jet gamma			&	$\gamma_j$		&	$4/3$				& \\
		\hline
		jet density			&	$\rho_j$		&$1.73\times10^{-26}$	& g/cm$^3$\\
		\hline
		jet nozzle radius	&	$r_{\text{jet}}=r_2$& $7.5\times10^{20}$	& cm\\
							&					& 243					& pc\\
		\hline
		jet magnetic field	&	$B$				& $1.7\times10^{-4}$	& gauss\\
		\hline
		jet plasma beta		&	$\beta_p$		& 1						& \\
		\hline
		jet internal Mach number & $M_j$		& 10					& \\
		\hline
		\parbox[t]{3.2cm}{jet toroidal to poloidal \\field ratio}	& $h$	& $\infty$, 1, 0	& \\
		\hline
		mean molecular weight	& $\mu$			& 0.61					& \\
		\hline
		ICM core density & $\rho_0$				& $9.6\times10^{-26}$	& g/cm$^3$\\
		\hline
		ICM density profile beta & $\beta$		& 0.53					& \\
		\hline
		\parbox[t]{3.2cm}{ICM density profile \\core radius} & $r_c$	& 26		& kpc\\
		\hline
		ICM core temperature&$T_\text{core}$	&3.0					& keV\\
							&					&$3.5 \times 10^7$		& K\\
		\hline
		ICM outer temperature &$T_\text{out}$	&6.4					& keV\\
							  &					&$7.4 \times 10^7$		& K\\
		\hline
		\parbox[t]{3.2cm}{ICM temperature profile\\ core radius} & $r_\text{c,T}$	& 60	& kpc\\
		\hline
		ICM gamma		&	$\gamma_\text{ICM}$	&	$5/3$				& \\
		\hline
	\end{tabular}
\end{table}

\subsection{Initial Conditions}
\label{subsec:init_cond}

We simulate jet propagation into a non-magnetized cluster initially in hydrostatic equilibrium and inject highly magnetized ($\beta_p = p_{\text{gas}}/p_B \sim 1$) jets at the center of the cluster. Initial density and temperature profiles are set according to the observational constraints of the Perseus Cluster derived by \cite{zhuravleva:15}:

The density profile follows a spherically symmetric $\beta$-model:
\begin{equation}
	\rho(r) = \frac{\rho_0}{[1+(\frac{r}{r_c})^2]^{\frac{3}{2}\beta}}.
\end{equation}
The temperature profile is set such that the core of the cluster is cooler than the outskirts appropriate to cool core clusters like Perseus:
\begin{equation}
	T(r) = \frac{T_{\text{out}}[1+(\frac{r}{r_{\text{c,T}}})^3]}{\frac{T_{\text{out}}}{T_{\text{core}}}+(\frac{r}{r_{\text{c,T}}})^3}.
\end{equation}
The respective numerical value of each variable can be found in Table \ref{tab:variables}.

%The radial profiles of density, pressure, and temperature are plotted in Fig.~\ref{fig:ICM_profile}. 
The jet and the ICM fluids follow adiabatic equations of state with $\gamma_j = 4/3$ and $\gamma_\text{ICM} = 5/3$, respectively. The gravitational potential of the cluster, dominated by dark matter, is held constant throughout the simulation (thus eliminating the need to solve Poisson's equation for self-gravity).

%\begin{figure}
%\includegraphics[width=\columnwidth]{ICM_profile.pdf}
%\caption[Radial profiles of the ICM properties]{Radial profiles of the initial conditions implemented in %our simulations. The profile is constructed using the parameters inferred from X-ray observation of the %Perseus Cluster and the condition of hydrostatic equilibrium. Analytical prescriptions can be found in %Section \ref{subsec:init_cond}. The gravitational potential is set up such that the cluster is initially %in hydrostatic equilibrium.}
%\label{fig:ICM_profile}
%\end{figure}

\subsection{Injection of the Jets}
\label{sec:jet_injection}
Following \citet{heinz:06,morsony:10}, jets are injected into the grid through a cylindrical ``nozzle'' at the center of the simulation domain. The total energy output is governed by the choice of nozzle radius, velocity $v_j$, jet mass density $\rho_j$, Mach number $M_j$, and toroidal magnetic field strength $B_{\phi}$ inside the nozzle. The power of the jets $L$ can be expressed as the sum of the kinetic energy flux, the enthalpy flux, and the Poynting flux, which can be expressed as the integral over the faces of the cylindrical nozzle
\begin{equation}
\label{eq:jet_power}
	L = 2 \int v_j \left(\frac{1}{2}\rho_j v_j^2 + \frac{\gamma_j}{\gamma_j-1}P_j + \frac{B_{\phi}^2}{8\pi}\right) \cdot \dd {A},
\end{equation}
where the factor of 2 accounts for two faces of the nozzle. The velocity profile $v_{\rm z}(r)$ of the jet across the nozzle is constant within the nozzle radius ($r_2$ in Fig.~\ref{fig:B_v_profile_r}) and smoothly goes to zero (at $r_\text{out}$) in a tapering feather region set to be 1/4 of the nozzle radius for numerical stability:
\begin{equation}
	\label{eq:v_r}
	v_j(r) =
	\begin{dcases}
		v_{0} &\text{if } 0 < r \leq r_2\\
		\frac{v_{0}}{2}\left(1+\cos(\pi \frac{r-r_2}{r_{\text{out}}-r_2})\right)
		&\text{if } r_2 < r \leq r_{\text{out}}\\
		0 &\text{if } r > r_{\text{out}}
	\end{dcases}
\end{equation}

%Equation \ref{eq:jet_power} is used to calculate the total energy input throughout the nozzle. 
We use the jet's internal sonic Mach number $M_j=10$ to parameterize the ratio between density and pressure given a fixed fluid velocity
\begin{equation}
	M_j = \frac{v_j}{c_s} = v \sqrt{\frac{\rho_j}{\gamma_j P_j}}.
\end{equation}
Given that the jet is highly supersonic, the total flux is dominated by the kinetic energy term, with the other fluxes contributing only at the percent level.

The plasma $\beta_p$, or the ratio of thermal to magnetic pressure
\begin{equation}
	\beta_p = \frac{p_j}{p_B}
\end{equation}
defines the strength of the magnetic field inside the jet and is set to 1 for all magnetized jets. The magnetic pressure within the jet includes both toroidal ($B_{\phi}$) and poloidal ($B_z$) fields inside the nozzle, such that the magnetic pressure is
\begin{equation}
	p_B = \frac{B^2}{8\pi} = \frac{B_{\phi}^2 + B_z^2}{8\pi}
\end{equation}
Note that enforcing $M_j=10$ and $\beta_p=1$ to be identical for all MHD  simulations---a choice made to facilitate direct comparison---introduces very small (less than $\sim$1\%) variations in total jet power between simulations, which we consider immaterial.

The nozzle size is then set by the requirement that the jet be in net lateral pressure balance with its environment at injection. If this were not the case and the nozzle size were too large, strong re-collimation shocks would form, crushing the jet and quickly leading to numerical dissipation of some of the internal toroidal field, thus affecting our ability to control the internal field structure of the jet (if the jet nozzle were set too small, computational requirements would unnecessarily increase). The overall maximum resolution is set by our goal to resolve the jet with at least 16 cells across both transverse directions of the nozzle.

Together with the condition of $M=10$, the resolution constraint informs our choice of jet injection velocity, based on computational feasibility: While a speed as close to $v_{\rm j}=c$ would be ideal (even in non-relativistic simulations), the computational demand is proportional to the number of computational cells $N_{3D}$ and the Courant time step $\Delta t_{c}$, depending on resolution $\Delta R$ as $N_{3D} \propto \Delta R^{-3}$ and $\Delta t\propto \Delta R/v_{\rm j}$, respectively. The resolution itself depends on the injection velocity as $\Delta R \propto v_{\rm j}^{-1/2}$, and thus the total computation demand $N_{\rm compute} \propto N_{3D}\times N_{\rm step} \propto N_{3D}/\Delta t_{\rm c}$ depends on injection velocity as $N_{\rm compute} \propto v_{\rm j}^{3}$, in the worst-case scenario. Realistically, the number of cells in an AMR simulation grows less rapidly than described above, leading to a somewhat less dramatic dependence of $N_{\rm compute} \sim \ln{(\Delta R)} \Delta R^2/v \propto \ln{v_{\rm j}} \sim \Delta R^2/v_{\rm j} \propto v_{\rm j}^2$.

We thus chose $v=0.1c$ for computational feasibility, given the available resources. We discuss our mitigation strategies for this approximation further in \S\ref{sec:precession}. Clearly, future experiments at full relativistic speeds would be desirable once computationally feasible.

\subsection{Magnetic Field Injection}

We set up 3 different cases of magnetized jets with toroidal, helical, and poloidal fields in the jets (as depicted in Fig.~\ref{fig:mag_drawing}), along with a non-magnetized case (hereafter referred to as the hydro case) as a control case.
\begin{figure}
	\includegraphics[width=\columnwidth]{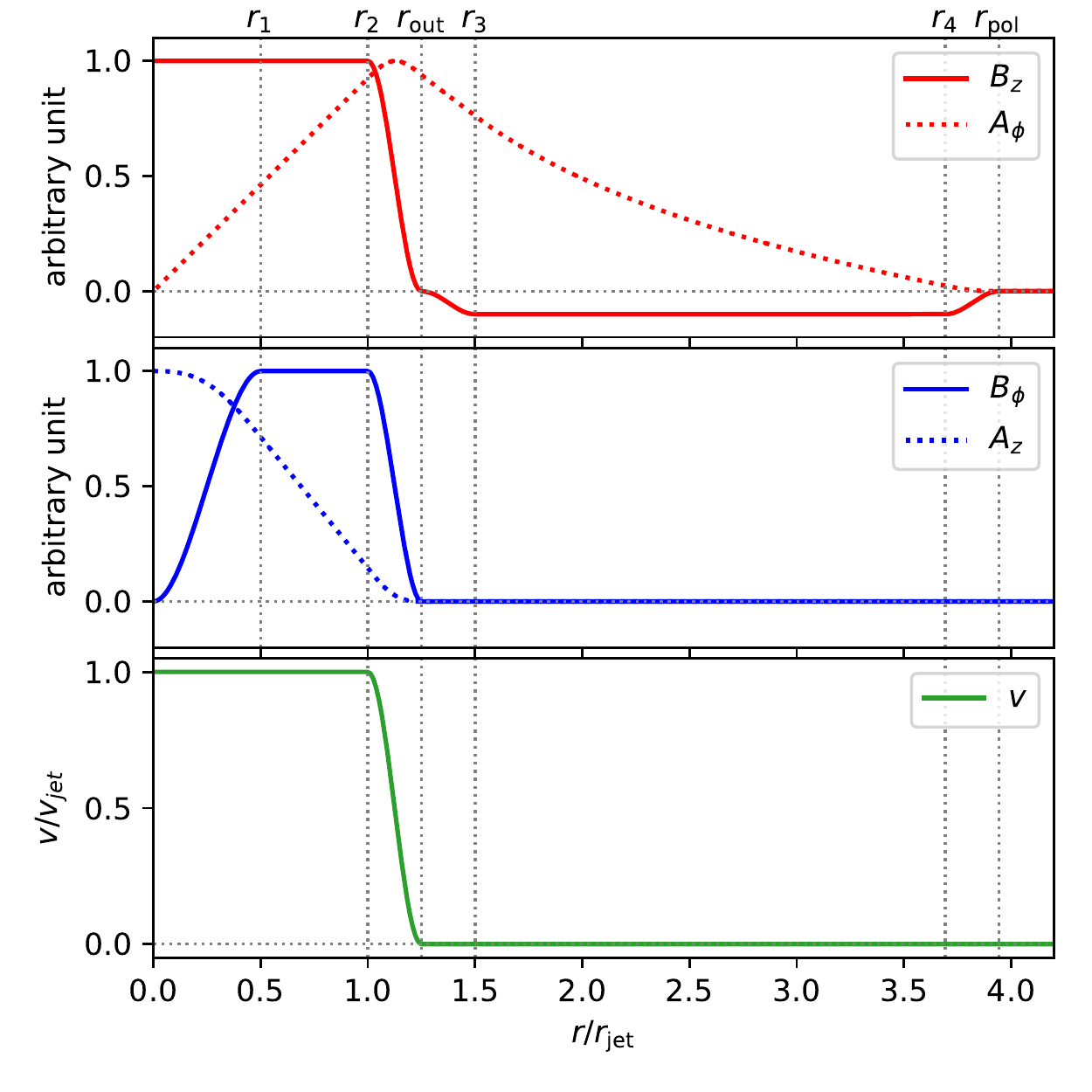}
	\caption[Radial profiles of the vector potential, magnetic fields, and
		velocity of the nozzle]
        {{{Radial profiles}} of the magnetic fields (upper and middle
		panel) and velocity (lower panel) inside and outside the nozzle. The
		inner tapering radius $r_1$ is set to be 1/2 of the nozzle radius. The tapering feather ($r_{\text{out}}-r_2$, $r_3-r_{\text{out}}$, and
		$r_{\text{pol}-r_4}$ ) is set to be 1/4 of the nozzle radius. The
		poloidal field ($B_z$) closes outside of the nozzle, where we can see
		negative values. The $B_z$ outside of the nozzle is set to be 10\% of
		the field in the nozzle and $r_4$ is the corresponding radius to close
		the field loop.}
	\label{fig:B_v_profile_r}
\end{figure}

We use a polynomial expression for $B_{\phi}$ to construct a smooth toroidal magnetic field inside the jet, which is constant ($B_{\phi 0}$) in most parts of the jet and approaches zero at the axis $r=0$ (to avoid diverging hoop stresses) and the boundary $r=r_{\text{jet}}=r_{\rm out}$. We split $B_{\phi}$ into 3 regions: 0 to $r_1$, $r_1$ to $r_2$, and $r_2$ to $r_{\text{out}}=r_{\rm jet}$. The polynomials are constructed such that the derivatives and thus B-fields are continuous at the boundaries. The following functions are used to set up the toroidal field:
\begin{equation}
	\label{eq:B_phi}
	B_{\phi} = B_{\phi 0}
	\begin{dcases}
		-\frac{2r^3}{r_1^3} + \frac{3r^2}{r_1^2} &\text{if } 0 < r < r_1\\
		1 &\text{if } r_1 < r < r_2\\
		-\frac{2(r_{\text{out}}-r)^3}{(r_{\text{out}}-r_2)^3} + \frac{3(r_{\text{out}}-r)^2}{(r_{\text{out}}-r_2)^2} &\text{if } r_2 < r < r_{\text{out}}
	\end{dcases}
\end{equation}

The poloidal field $B_z$ is set to be constant ($B_{z0}$) inside the jet and closed on field lines outside the nozzle such that the magnetic loop is closed. The initial $B_z$ outside of the nozzle is set at 10\% of the strength inside a region that guarantees zero net flux. We construct the functional form of $B_z$ to ensure smooth derivatives at the transitions. The resulting profile of the magnetic fields is shown in Fig.~\ref{fig:B_v_profile_r}.

\begin{figure}
	\includegraphics[width=\columnwidth]{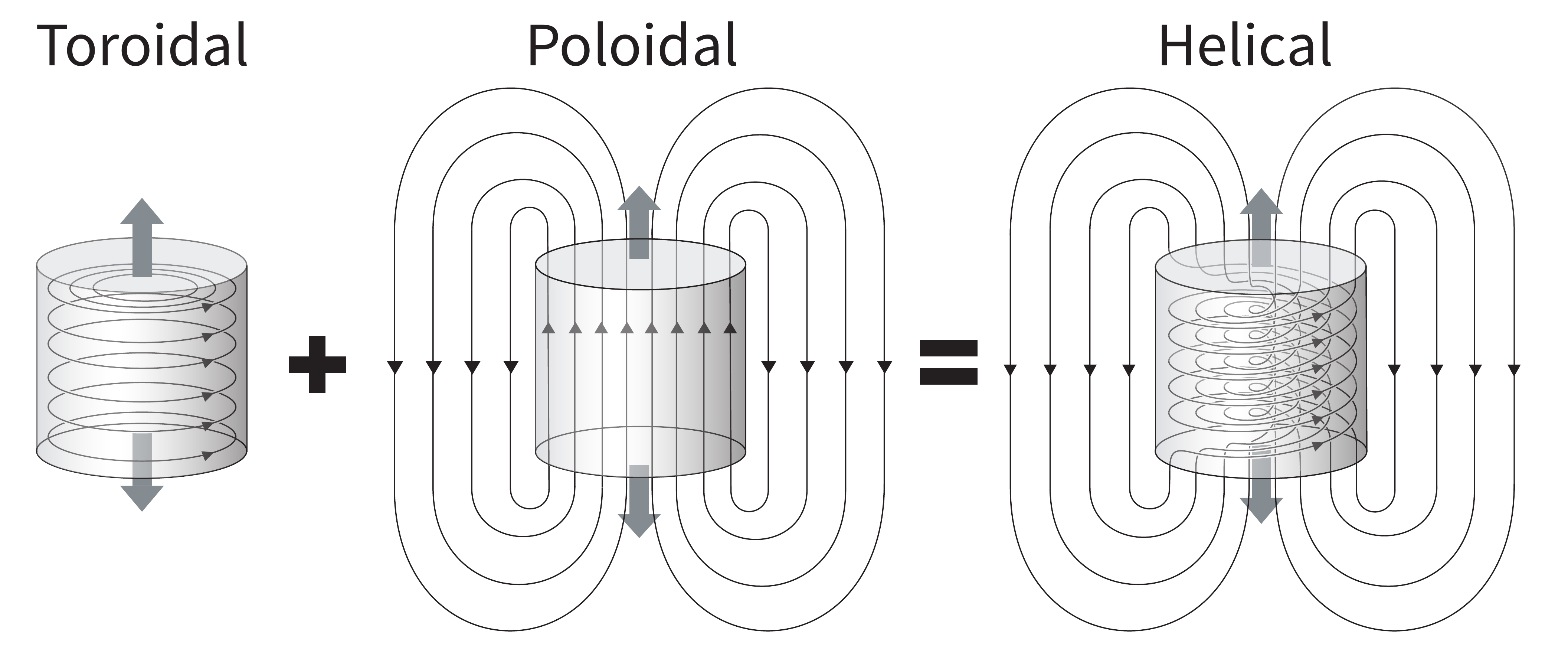}
	\caption[Schematic drawings of magnetic fields]{{{Schematic drawings}} of
		the injection magnetic topology of the jet: toroidal (\emph{left}), poloidal (\emph{middle}), and helical (\emph{right}), which is a weighted sum of the other two cases. The total magnetic field strength inside the region $r_{1}\leq r \leq r_{2}$ is set to be the same for all 3 cases. The shaded regions indicate the size of the jet nozzle. Note that the cartoon shows the poloidal field in the plane of paper, but the actual field is axi-symmetric.}
	\label{fig:mag_drawing}
\end{figure}

In practice, we use the magnetic vector potential $\mathbf{A}$ to prescribe the magnetic field to ensure we maintain a divergence-free field.
\begin{subequations}
	\begin{align}
	A_r =& 0\\
	A_{\phi} =& \frac{1}{r}\int r B_{z} \dd r\\
	A_z =& \int B_{\phi} \dd r
\end{align}
\end{subequations}

In order to replenish the toroidal field as the fluid moves out of the nozzle, we calculate the advection rate and add the appropriate amount of flux to maintain the same magnetic field strength throughout the active period of the jet. This field injection region is downstream of the hydrodynamic nozzle (since the fluid properties inside the nozzle are not updated but instead serve as an interior inflow boundary condition).

The helicity parameter $h$ is used to control the ratio between toroidal and poloidal components inside the jet nozzle
\begin{equation}
	h = \frac{B_{\phi0}}{B_{z0}}.
\end{equation}
We set $h=\infty$ for toroidal case, $h=1$ for helical case, and
$h=0$ for poloidal case. The total magnetic field pressure is set to be the
same for all 3 cases, such that
\begin{equation}
	(B_{\phi}^{\text{toroidal}})^2
	= (B_{\phi}^{\text{helical}})^2 + (B_{z}^{\text{helical}})^2
	= (B_z^{\text{poloidal}})^2
\end{equation}

\subsection{Jet Jitter}
\label{sec:precession}
We include a modest, random variation in the jet direction around the mean jet axis, which is oriented along the $z$-axis of the simulation box. This jitter is imposed for two reasons:

First, it is known from observations that jet axes change direction on timescales comparable to the jet travel time \citep[e.g.][]{nulsen:02,young:05}. This may be due to bona fide jet precession, but even in the absence of an ordered, secular variation in jet axis, dynamical instabilities in the accretion disks and at the base of the jets are likely to cause moderate changes in orientation \citep{pringle:96,heinz:00}. Extreme re-orientation between episodes of jet activities could even result in X-shaped radio sources that have been observed \citep{dennett-thorpe:02,roberts:18}. This, however, is not what we are modeling in this work.

Second, analytic and numerical models of uni-directional jets do not generally reproduce the well-known shape of radio galaxies, leading to much larger aspect ratios of cocoons and radio lobes than observed \citep[e.g.][]{norman:82,marti:94}, insufficient to couple the AGN energy to the central cluster \citep{vernaleo:06}. This led \cite{scheuer:74,scheuer:82} to propose the so-called dentist drill effect, which spreads jet thrust over a much larger cone than that prescribed by the jet's opening angle. Several numerical studies also show that precession of the jets is necessary to match the observed morphology of the radio emission \citep[see e.g.][]{heinz:06,nawaz:16}.  

\begin{figure}
\includegraphics[width=\columnwidth]{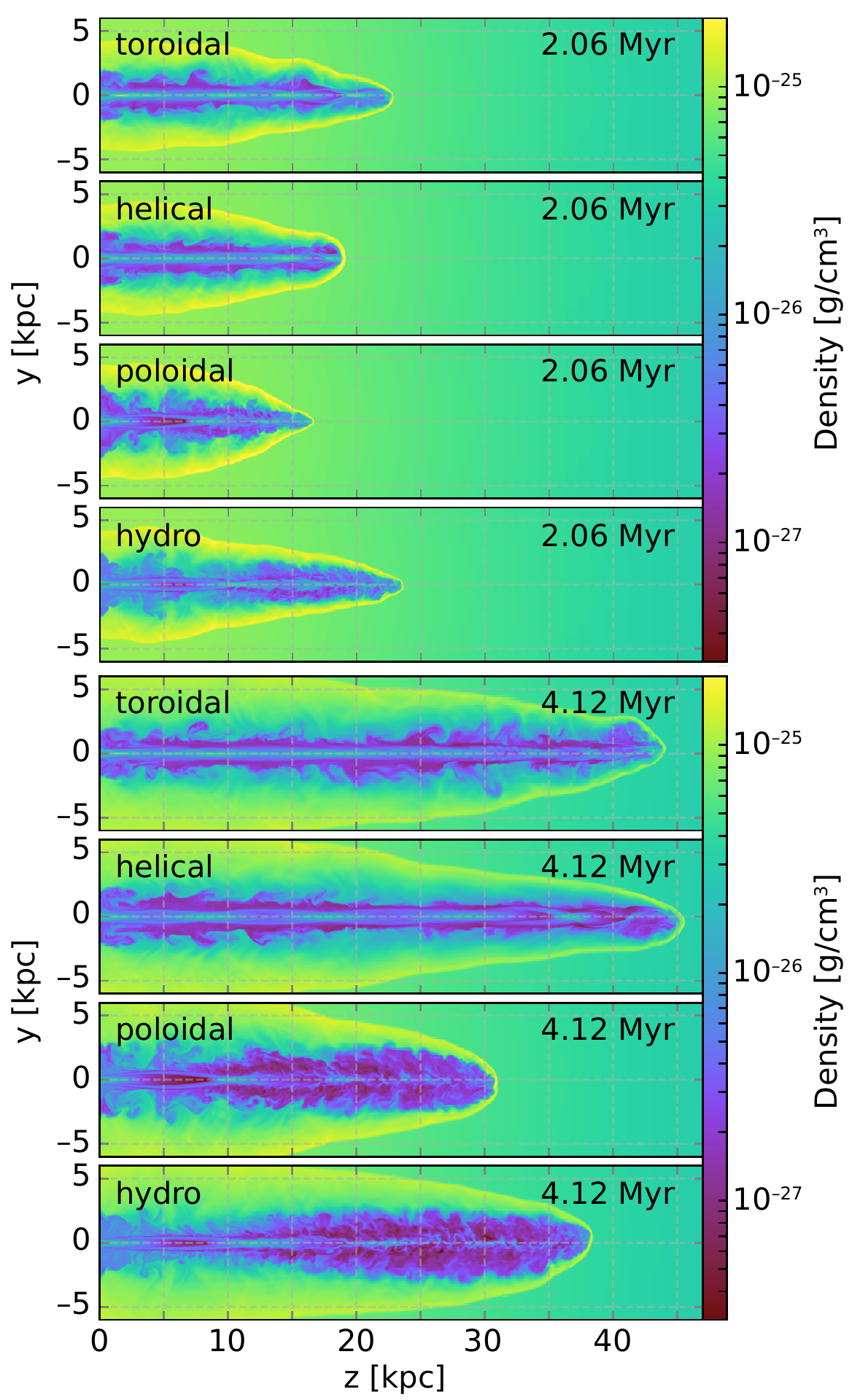}
\caption[Density slices of non-precessing jets]{Slices showing density through the central plane in simulations with constant jet direction (non-precessing) at {{2.06 and 4.12 Myrs (top and bottom four panels, respectively). In each age group, from top to bottom: pure toroidal, helical, poloidal, no magnetic field.}} One can see the effect of the magnetic field on the propagation of jets. The toroidal component of the magnetic field imposes hoop stress on the flow and helps collimate the jets, while the poloidal component increases the lateral pressure and makes the jet expand, but stabilizes it against kink instability. See text for more discussion.}
\label{fig:compare_4_nojiggle}
\end{figure}

This is particularly important in simulations like ours that are not fully relativistic: Relativistic jets with a velocity close to $c$ have an energy-to-momentum ratio of about $c$, while in our simulation the speed of the jets is set to be 0.1 $c$ and has a correspondingly smaller energy-to-momentum ratio. The excess momentum leads to an  elongated shape that does not fit observations (see Fig.~\ref{fig:compare_4_nojiggle}), but can be remedied by distributing this momentum over a larger solid angle.

\begin{figure}
\includegraphics[width=\columnwidth]{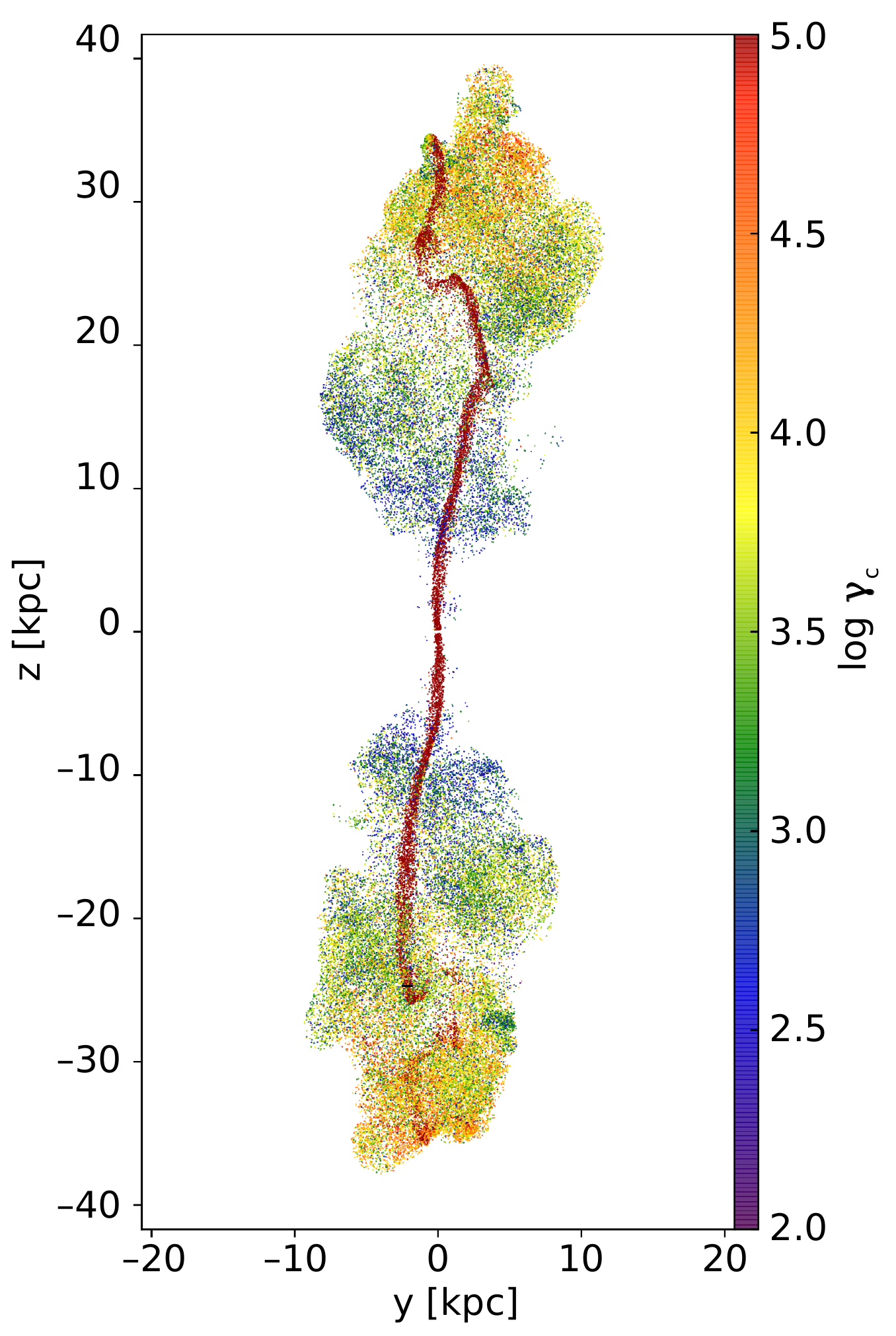}
\caption[Projection of particle $\gamma_{\text{cut}}$]
	{Projection of particle $\gamma_{\text{cut}}$, which corresponds to the
	cut-off energy of the electron distribution, at 10 Myr. Each particle,
	which represents an ensemble of electrons, is accelerated in the
	termination shocks, where it starts from a power-law distribution with
	infinitely large cut-off energy. As particles propagate with the fluid,
	their cut-off energy is changed due to synchrotron radiation and adiabatic
	expansion/compression.}
\label{fig:gamc}
\end{figure}

\begin{figure}
  \includegraphics[width=\columnwidth]{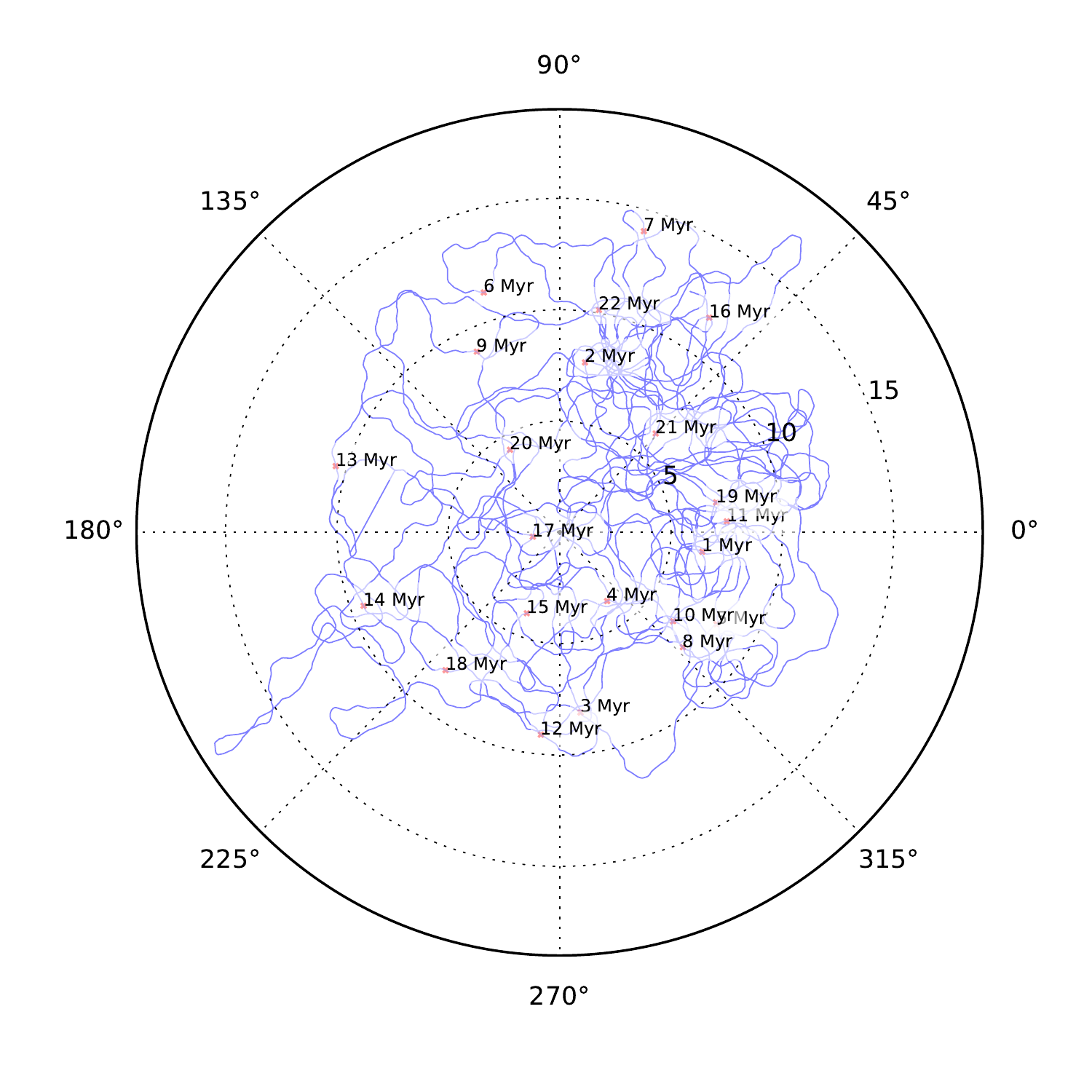}
  \caption[Pointing of the jet directions]{The pattern of the jet pointing direction as implemented in the simulation. The pointing undergoes a random walk process but is constrained by a maximal angle that can be set in the simulation. Red crosses label the position and time of the pointing since the launch of the jet. The labels on the radial-axis are in units of degrees, from 0 to 20$^{\circ}$.}
  \label{fig:jiggle_h1}
\end{figure}

We implement this effect by allowing the direction of the jet to drift in a random walk prescription, limited within a set opening angle. One can see the jets are bent in Fig.~\ref{fig:gamc} due to the introduction of this jitter. The resulting motion of the jet axis throughout the simulation is shown in Fig.~\ref{fig:jiggle_h1}. The opening angle of this cone is modest---the average angle between the jet and the $z$-axis is about 8 degrees, {{broadly similar to the typical change one might expect due to  dynamical instabilities within the jet (as explored in a rich body of literature on dynamical stability of relativistic jets; see, e.g., \citealt{bodo:19} and references therein) and potential changes in jet direction from either disk precession or change in the net angular momentum of accreted material \citep{bicknell:96,caproni:04,erlund:07,marti-vidal:11,osullivan:12,babul:13,britzen:18,dominik:21} and what is required to explain observed aspect ratios of radio lobes \citep{scheuer:82,heinz:06}}}.

{{It is important to point out that the jitter pattern is fully randomized between simulations analyzed in this paper, i.e., each simulation with jitter follows a statistically similar but different random pattern. We also performed a set of simulations with identical dither patters (otherwise following the same range of magnetic topologies) to verify that our results are robust. We find that material differences arise only during the first 3 Myrs.}}

To isolate effects unrelated to this jitter in our simulations, we also performed a control set of simulations of non-precessing jets with otherwise identical parameters, but run for a shorter time of 5Myrs. Results from these control runs are presented in \S\ref{sec:nojiggle} and \S\ref{sec:kink_instability}. Unless otherwise mentioned, our discussion throughout the paper focuses on the set of simulations including jitter.

\subsection{Solving the Transport Equation for Tracer Particles}
\label{subsec:tracer_particles}

The synchrotron intensity of a synthetic radio map generated from a simulation frame depends on (a) the number density of electrons (and potentially positrons), which is derived from the energy density in the simulations, assuming the internal energy of the non-thermal plasma is dominated by the relativistic leptons, and (b) the strength and orientation of the magnetic fields.  To obtain the spectra of the synchrotron emission, one further needs an estimate of (c) the energy distribution of the electrons. There are several ways to account for (c), including assuming a simple power-law \citep[e.g.][]{hardcastle:14,english:16}, grid-based electron population transport model, for example in \cite{del-zanna:06,mendygral:12}, or Lagrangian particle-based electron transport in \citep{mimica:09,vaidya:16,vaidya:18}.

To this end, we implement a simple yet accurate numerical framework to calculate synchrotron emission in our simulations, employing the Lagrangian tracer particle framework in \code{FLASH} \citep{dubey:12}, whereby the  radiative and adiabatic expansion/compression losses/gains of relativistic electrons/positrons are tracked by individual computational Lagrangian tracer \emph{particles} (as opposed to the Eulerian jet tracer {\em fluid} that tracks the fraction of jet fluid in each cell, which our simulations also employ). Note that the term \emph{particle} here refers to the numerical perspective of a Lagrangian fluid element that moves spatially during the simulation. Each \emph{particle} represents an ensemble of non-thermal leptons that are modeled in the spatial proximity of the location of the tracer particle.

Following \citet{bicknell:82,coleman:88}, we implement the well-known characteristic solution to the particle transport equations subject to radiative losses and adiabatic compression. 

We assume the synchrotron radiation from the radio lobes is optically thin to self-absorption, and the radiation energy is negligible compared to the thermal energy of the gas. This is consistent with most extended radio sources observed, which have steep spectral indices \citep[see e.g.][]{laing:83}.

Particles are injected at the cross-sections of the jet nozzle and are assumed to initially represent a power-law distribution of relativistic electrons
\begin{equation}
    f(\gamma,t=t_{0})=N_{0}\gamma^{-p}
\end{equation}
for $\gamma \geq \gamma_{\rm min} = 10$. However, any injection spectrum can in principle be modeled this way in post-processing.

While $p$ is a free parameter, we choose $p=2$ throughout, given the optically thin synchrotron spectral index of $\alpha = -d\log{F_{\nu}}/d\log{nu} = (p-1)/2 \sim 0.5$ for typical young radio sources.  The normalization $N_{0}$ is then related to the non-thermal particle pressure $P$, as
\begin{equation}
    N_{0}=\frac{3P}{m_{\rm e}c^{3}\ln{\gamma_{\rm max}/\gamma_{\rm min}}}
    \label{eq:normalization}
\end{equation}
Given the very weak logarithmic dependence on the limits of the normalization on $\gamma_{\rm min}$ and $\gamma_{\rm max}$, we set $\gamma_{\rm min}=10$, while we track the evolution of $\gamma_{\rm max}$ as described below.

For each \emph{particle}, the mapping of the initial distribution to the time-evolved distribution is accomplished using the characteristic solution for particles injected with energy $\gamma_{0}$ as a function of time:
\begin{equation}
	\label{eq:gamma_crit}
	\gamma_{\text{c}}(\gamma_{0},t) = \frac{\left( \frac{n(t)}{n_0} \right)^{\frac{1}{3}}}
	{1/\gamma_{0} + \int_{t_0}^t A(t') \,(\frac{n(t')}{n_0})^{\frac{1}{3}}\dd t'}
\end{equation}
where $n(t)$ is the number density of electrons at time $t$, $n_0$ is the
initial number density of \emph{leptons}, and $A = \frac{4 \sigma_T}{3 m_e c} \left(U_B + U_{\rm CMB}\right)$ is the radiative energy loss rate, neglecting synchrotron-self-Compton losses (which are insignificant for electrons in diffuse synchrotron emitting regions like radio lobes). For an injection spectrum of $p=2$, the solution reduces to a simple power-law with the same index but a sharp cutoff at
\begin{equation}
    \label{eq:gamma_cut}
    \gamma_{\rm cut}=\lim_{\gamma_{0}\rightarrow \infty}\gamma_{c}(\gamma_{0},t)=\frac{\left( \frac{n(t)}{n_0} \right)^{\frac{1}{3}}}
	{\int_{t_0}^t A(t') \,(\frac{n(t')}{n_0})^{\frac{1}{3}}\dd t'}  
\end{equation}

In the simulations, we have to choose a time $t_{0}$ to begin the integration, i.e., a reference time when the electrons are accelerated to have a power-law energy distribution.  The acceleration mechanism of particles inside jets is still not fully understood; the site of the acceleration could be located inside the jet \citep{bicknell:82} and/or at terminal hotspots, and the mechanism could be, e.g., magnetic reconnection \citep{romanova:92,sironi:14,sironi:15,guo:15} or diffusive shock acceleration \citep{bell:78,blandford:78,drury:83,caprioli:14,park:15}. 

We choose to begin the cooling calculation when the fluid element the particle travels in has dissipated most of its kinetic energy, identified as the time at which the bulk velocity of the \emph{particles} first drops below half of the jet injection velocity.

This approach requires significantly less computational effort and memory than solving the full transport equations numerically \citep{mendygral:12,vaidya:18}. More importantly, the existence of a closed, analytic solution to the transport equation makes this approach more accurate than a finite-difference solution of the same equation.

\subsection{Synchrotron Emissivity}
\label{subsec:synchrotron_emis}

We map the synchrotron spectrum of individual tracer particles onto the grid and produce synthesized radio observations of the simulated data in post-processing by projecting along a given line-of-sight. The synchrotron spectra of the grid cells are determined following the standard expressions for the integrated synchrotron emission coefficient \citep[e.g.][]{rybicki:79} for an isotropic pitch angle distribution.

Given the sharp cutoff to the particle distribution the shape of the synchrotron spectrum for each tracer particle is well-described by a simple power-law with an exponential cutoff, such that the emission coefficient in each cell is 
\begin{equation}
    j_{\nu} 
    \approx \frac{3\left(B\sin{\alpha}\right)^{3/2}e^{7/2}N_{0}}{8\sqrt{\pi}c^{5/2}m_{\rm e}^{3/2}}\nu^{-1/2}C_{0}e^{-\left(\nu/\nu_{\rm max}\right)}
	\label{eq:Jnu_final}
\end{equation}
with $\alpha$ being the angle between the magnetic field vector and the line of sight such that $B_{\perp}=B\sin{\alpha}$ is the in-sky component of the B-field, and, for $p=2$,
\begin{equation}
    C_{0}=\frac{4\sqrt{2}}{3}\Gamma \left(2 \frac{1}{12}\right)\Gamma\left(\frac{5}{12}\right)\approx 4.16
\end{equation}
The cutoff frequency is
\begin{equation}
    \nu_{\rm max}=\frac{3\gamma_{\rm max}^2 e B\sin{\alpha}}{4\pi m_{\rm e}c}
\end{equation}
in which $\gamma_{\text{max}}$ is evaluated for each tracer particle using eq.~(\ref{eq:gamma_cut}); the normalization $N_{0}$ is calculated using eq.~(\ref{eq:normalization}); and pressure $P$ and magnetic
field $B$ are taken from the simulation grid. 

In practice, we calculate the value of $j_{\nu}$ in each cell through a weighted average of the nearest 10 particles, weighted by inverse square distance from the cell center. This is necessary because not every computational cell contains at least one tracer particle.

The emission coefficient can then be integrated along the line-of-sight to calculate the synthesized synchrotron intensity maps, which are presented in Section \ref{subsec:synchrotron_properties}. Given the fact that the emission clock of each tracer {\em particle} is started after the jet fluid has dissipated its kinetic energy, we remove any emission from the jet fluid itself (i.e., removing any computational cell with velocity in excess of half the jet speed) and focus our analysis of the synchrotron emission on the radio lobes.

\section{Results}
\label{sec:results}

A motivating goal of our simulations is to investigate how different topology of the magnetic fields in the jets affect the jet-ambient medium interaction, including the collimation, propagation, and the imprints on the radio lobes, and to probe which physical effect is responsible for any observed differences. We thus compare the outcome of simulations with different injected magnetic topology to each other and to a purely hydrodynamic control simulation.

\subsection{Non-Precessing Jets}
\label{sec:nojiggle}
We perform a set of control simulations without explicit precession to serve as our calibration and baseline cases. These non-precessing simulations have the same configuration as our precessing simulations, but without the jitter described in \S\ref{sec:precession}. The simulations were run for a shorter duration of 5 Myrs.

Snapshots of density slices are shown in Fig.~\ref{fig:compare_4_nojiggle}. We can see that each jet propagates at different speed into the external medium. Although the jet velocity is the same across all simulations, the head of the jet is influenced by the internal dynamics of magnetic fields and thus has different ability to propagate.

We determine the propagation velocity of the jet head in the simulations by measuring the endpoint of the jet tracer fluid in the direction of the jet axis (z-axis). This is achieved by creating profiles of the jet fluid mass fraction against the z-coordinate. We then determine the location of the jet head to be where this fraction $f_{\rm W}$ exceeds $10^{-6}$. We analyze both sides (+z and -z) and the difference between them can be considered as the numerical uncertainty for the analysis.

\begin{figure}
\includegraphics[width=\columnwidth]{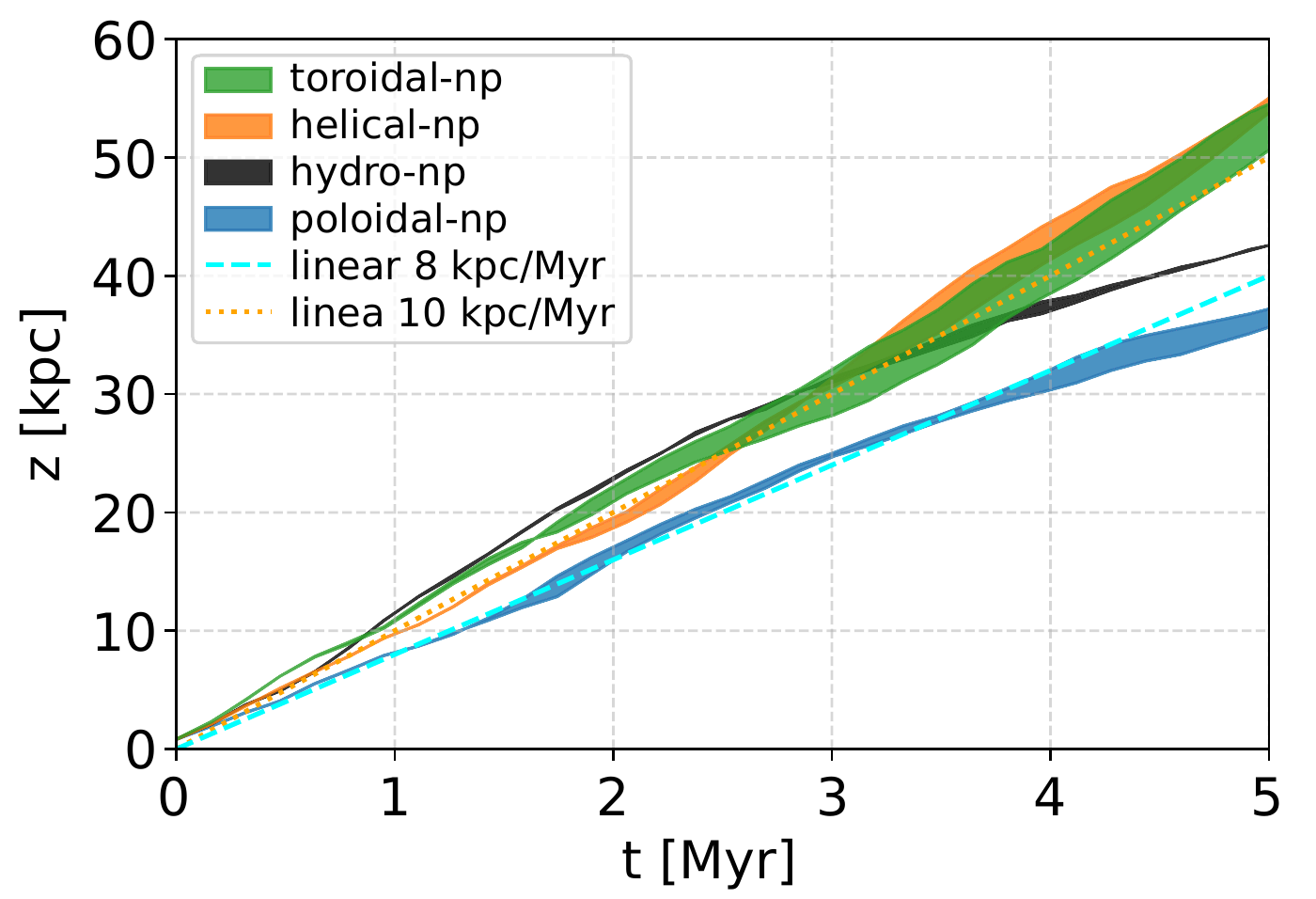}
\caption[Propagation distances of the jet heads for the 4 non-precessing jets]{Propagation distances of the jet heads for the 4 non-precessing jets. The edges of the filled region represent the longest extension of the jet fluid in the +z and -z directions for simulations without precession. Dotted and dashed lines: linear propagation rate for comparison. Note that the jet fluid is set to a velocity of 0.1 $c$ $\sim$ 30 kpc/Myr in our simulations.}
\label{fig:lobe_size_nojiggle}
\end{figure}

Fig.~\ref{fig:lobe_size_nojiggle} shows the results of this analysis. The edges of the shaded regions represent the location of the jet heads. We find that the toroidal jet propagates the fastest at the beginning (before 1 Myr) followed by the helical and hydro jets, while the poloidal jet is the slowest. Later, the hydro jet becomes slightly faster than all others between 1 and 3 Myr, but slows down after about 3 Myr. Finally, after 4 Myr, the poloidal jet remains the slowest, while the hydro jet becomes slower than the helical and toroidal jets. The hydro jet propagates at a velocity comparable to the poloidal jet.

\begin{figure}
\includegraphics[width=\columnwidth]{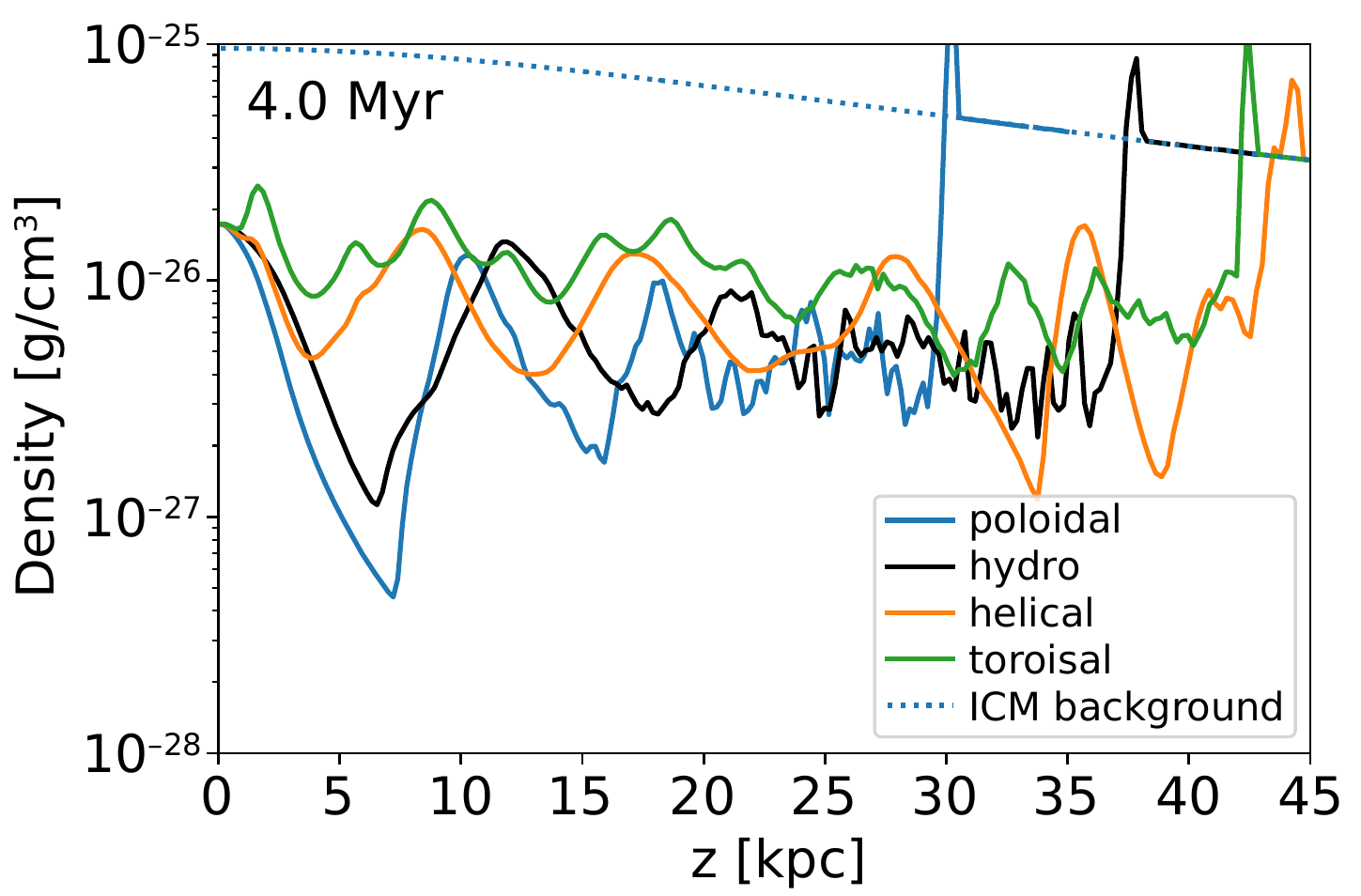}
\caption[Averaged density profile of the non-precessing jets at 4 Myr]{Averaged density profile within 0.25 kpc from the axis of the non-precessing jets at 4 Myr. All simulations begin with the same density in the nozzle, but the different configurations of the magnetic fields give rise to different density profiles after the jet fluid leaves the nozzle. This causes the different propagation velocity of the jet heads in our four cases.}
\label{fig:line_profile_nojiggle}
\end{figure}

Analytically, the propagation velocity of the jet head can be estimated by
balancing the ram pressure in the frame of the jet head once the density ratio and the bulk velocity in the jets is known \citep[see e.g.][]{norman:85}. The velocity of the jet head can be written as
\begin{equation}
	v_h = \frac{\sqrt{\eta}}{1+\sqrt{\eta}}v_j,
\end{equation}
where $v_j$ is the jet fluid velocity and $\eta = \rho_j / \rho_e$ is the ratio of the jet fluid density and environmental density. In our cases, the jet velocity is fixed to be 0.1 $c$ ($\sim$ 30 kpc/Myr), and the environmental density profile is the same for all simulations. The variation of the propagation velocity thus arises from the density of the jet. Although the density of the nozzle is the same for all simulations, the different configurations of magnetic fields cause the jet to expand or self-collimate at different rates. Fig.~\ref{fig:line_profile_nojiggle} shows the density profile along the jet axis at about 4 Myr.  Two reference lines of constant head velocity $v_h$ are plotted in Fig.~\ref{fig:lobe_size_nojiggle}: 8 and 10 kpc/Myr for density ratio $\eta$ of 0.12 and 0.23 respectively.

Even though in all simulations the jets begin with the same density in the nozzle, the magnetic structure determines the later evolution of the jet. The figure shows that the jet in the toroidal case (green line) maintains the highest mean density as it propagates, while the jet in the poloidal case drops to very low density as soon as the fluid leaves the nozzle. 

Based purely on the lateral force balance of a steady jet, one would thus expect that the toroidal jet should propagate fastest due to the hoop stresses imposed by the toroidal magnetic fields on the flow, which collimate the jet, compress the flow, and increase the density, leading to a higher momentum flux, thus enabling the fluid to penetrate faster in the ICM.  While this is true initially, we find that the toroidal jet slows down to speeds below that of the helical case.

On the other hand, the poloidal magnetic fields act as additional lateral pressure that expands the jet and reduces the density, making it slower to propagate, which suggests that the poloidal case should have the slowest propagation velocity.

However, the high density and degree of collimation of the toroidal case do not consistently make the toroidal jet propagate the fastest. As seen in Fig.~\ref{fig:lobe_size_nojiggle} and the right panel of Fig.~\ref{fig:compare_4_nojiggle}, the helical jet propagates as fast as the toroidal jet at a later time ($\gtrsim$ 3 Myr), despite the lower toroidal field and the added lateral pressure due to the poloidal field. The reasons for this are described in detail in Section \ref{sec:kink_instability}.

\subsection{Jet Propagation with Jitter}
\label{sec:dynamical_properties}

\begin{figure}
\includegraphics[width=\columnwidth]{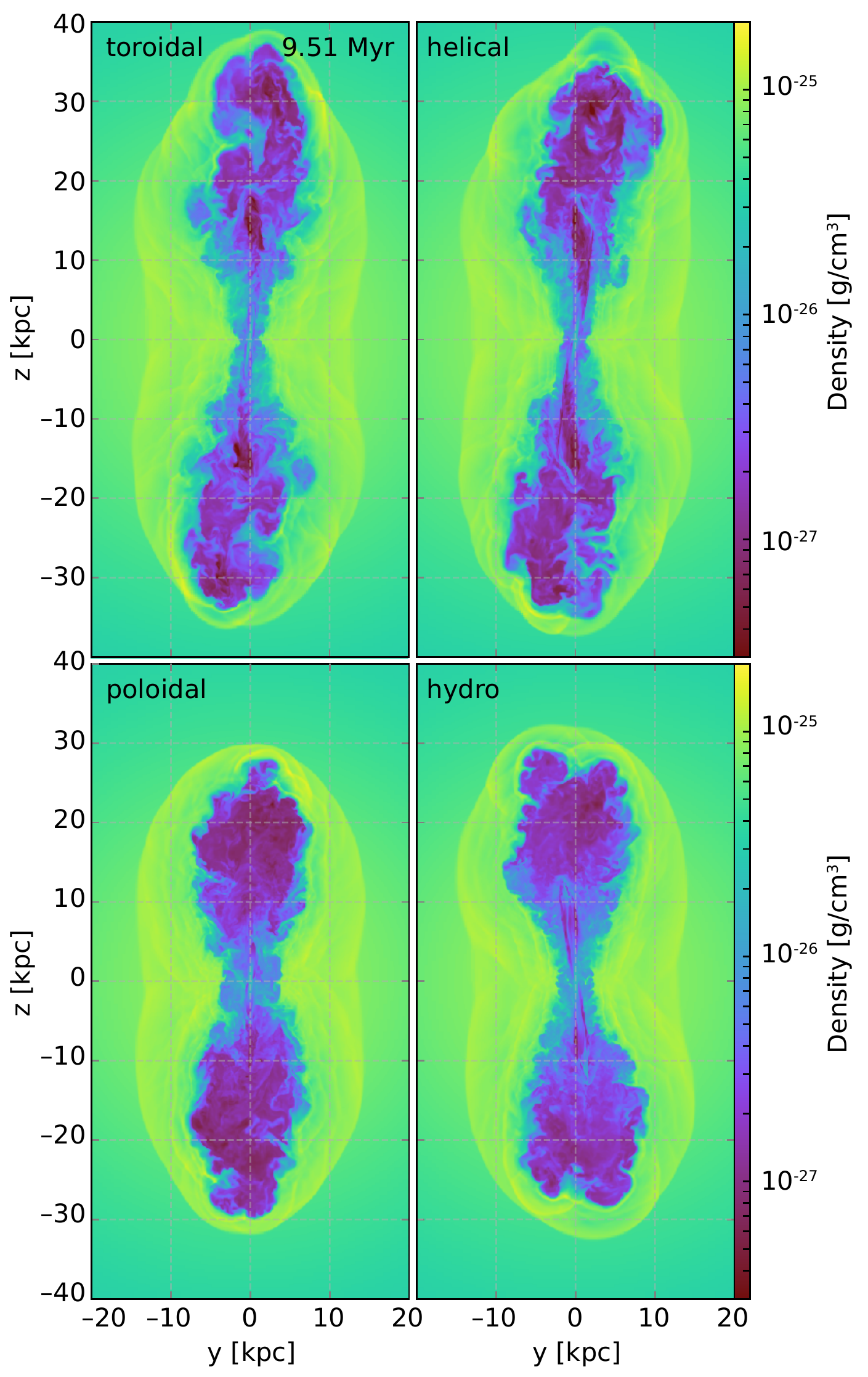}
\caption[Density slices of the simulations at 10 Myr]
29{Slices showing the density through the central plane from four simulations with different magnetic field topology in the jets at 10 Myr. {\em Top left:} pure toroidal, {\em top right:} helical, {\em bottom-left:} poloidal, {\em bottom-right:} no magnetic field. }
\label{fig:compare_4}
\end{figure}

When jitter is implemented (Section \ref{sec:precession}), the shapes of the cavities look very different from the non-precessing jets. The cavities created by the jets resemble the lobes of typical FR-II radio galaxies \citep[][; some typical examples include Cygnus A, Centaurus A, Fornax A and Hydra A]{fanaroff:74}, whereas in the case without jitter, the cavities/lobes are very skinny, unlike those of actual radio galaxies. In Fig.~\ref{fig:compare_4}, we show the central slices of the four simulations at about 10 Myr. In all cases, the jets have been continuously active for 10 Myr, after which the jet is turned off and the lobes are allowed to evolve passively for another 91
Myrs. 

The lobe shapes of the toroidal and helical cases are noticeably different from the poloidal and hydro cases. When toroidal fields are included in the jet (both toroidal and helical cases), the cavities are more elongated, with larger aspect ratios, compared to those without toroidal fields (poloidal and hydro cases), with narrower waists in the equatorial region of the radio galaxy.

\begin{figure}
\includegraphics[width=\columnwidth]{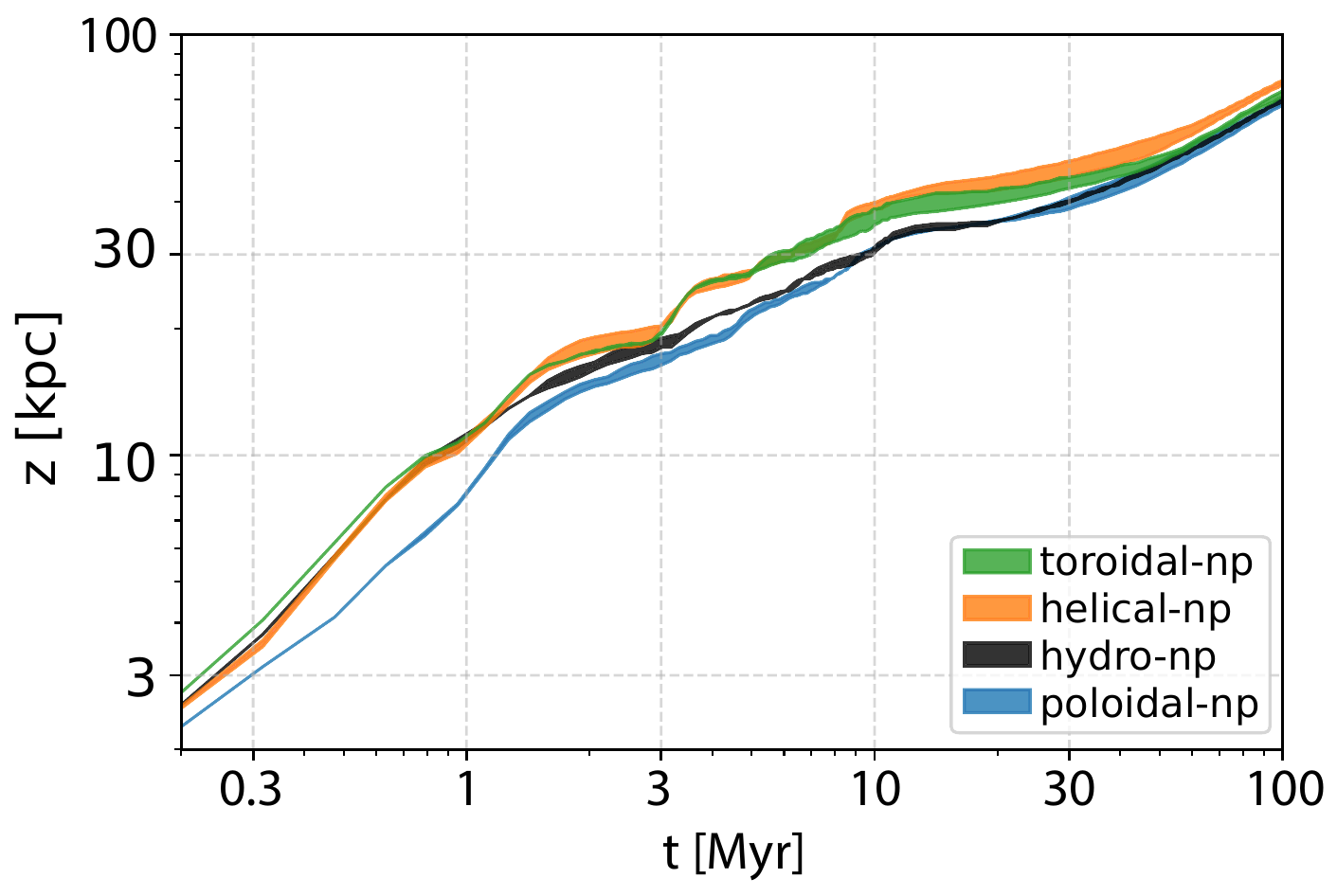}
\caption[Propagation distances of the jet heads]
	{Evolution of the lobe extent in the four precessing simulations. Edges
	of the shaded area represent the furthest extents of the jet fluid in +z
	and -z directions, representing two sides (upper and lower) of the jets and
	can be interpreted as numerical uncertainties in the simulations. The jets
	are turned off at 10 Myr. Note that both axes are in log-scale.}
\label{fig:lobe_size}
\end{figure}

To quantify the differences in propagation, we once again measure the extent of the lobes from our simulations, similar to the non-precessing case. The location of the jet heads is plotted in Fig.~\ref{fig:lobe_size}. At the very early stage ($<$ 1 Myr), we see the same sequence of toroidal $>$ hydro $\gtrsim$ helical $>$ poloidal in lobe sizes. However, after 1 Myr the toroidal case slows down to speeds comparable to the helical case, and the hydro case slows down to speeds comparable to the poloidal case. At 10 Myr, the time shown in Fig.~\ref{fig:compare_4}, they form two distinct groups with toroidal $\sim$ helical $>$ hydro $\sim$ toroidal. {{As pointed out in \S\ref{sec:precession}, the simulations with four different magnetic topologies follow different randomized jitter patterns. In order to verify that the differences in lobe propagation are unaffected by this randomization, we repeated this analysis using re-simulations with identical dither pattern. The only material difference between these two sets of simulations appear during the first 3 Myrs. Because our analysis of the simulations with jitter focuses on the later stages of the simulations, we conclude that this difference at early times does not affect our conclusions in any meaningful way.}}

The behavior can be understood in the context of the non-precessing control runs in Section \ref{sec:nojiggle}. The toroidal and helical cases, given the action of magnetic hoop stress, can collimate better and carry more ram pressure during the propagation, resulting in the cavities being more elongated. On the other hand, the poloidal and hydro cases lack any internal collimation mechanism and propagate more slowly, after an initial period during which lateral pressure balance is established.

However, the pure toroidal case once again does not propagate faster than the helical case. This is in contradiction to the understanding that the propagation velocity is determined by the collimation of the jet and thus by the strength of toroidal fields. We have seen similar results in the non-precessing case, in which the propagation of the pure toroidal case slows down to be comparable to the helical case. 

Similarly, the hydrodynamic case approaches the propagation speed of the poloidal field case, despite the absence of the additional lateral pressure resulting from the poloidal field. We attribute these deviations from the naive expectations to the development of dynamical instabilities, which we will discuss in \S \ref{sec:kink_instability}.

\subsection{Synchrotron Properties}
\label{subsec:synchrotron_properties}

\begin{figure}
\includegraphics[width=\columnwidth]{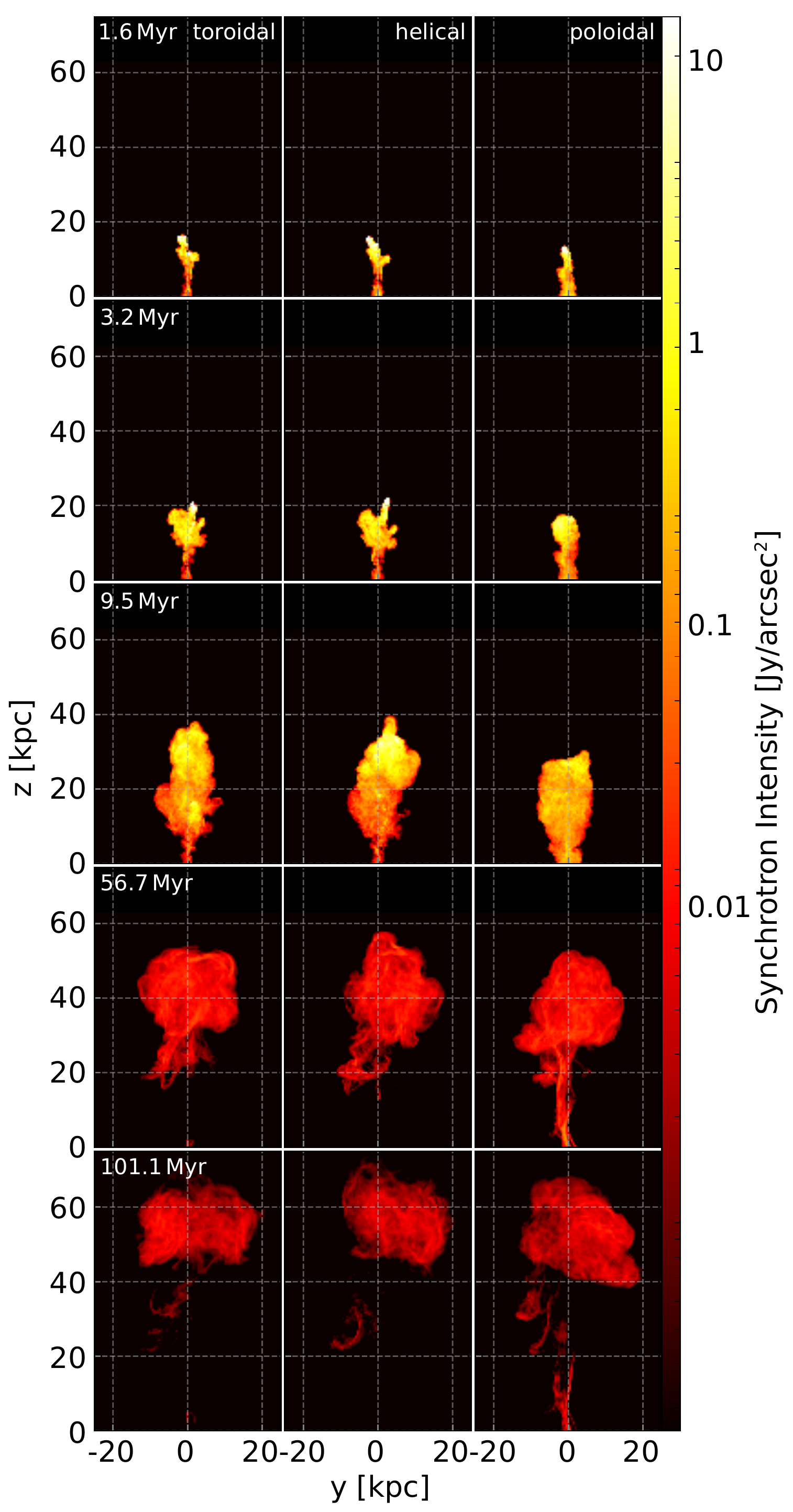}
\caption[Time series of the synthesized synchrotron maps at different times]{Evolution of synchrotron emission at different times in the simulations as seen at 90 degree inclination of viewing angle to jet axis. \emph{Left} column: toroidal magnetic fields; \emph{middle} column; helical magnetic fields; \emph{right} column: poloidal magnetic fields.}
\label{fig:sync_150MHz}
\end{figure}

With the tracer particle implementation outlined above, we can synthesize the synchrotron radiation at various frequencies of different evolutionary stages for different magnetic injection topologies. We show a sequencce of synchrotron emission maps at 150 MHz in Fig.~\ref{fig:sync_150MHz}. In the early stages, due to the precession of the jets, the shape of the radio image is spiky with various episodes of the jet rapidly propagating into the ambient medium as jitter changes the jet direction. The emission is brightest at the endpoints of the spikes, where the jet fluid runs into the denser medium and decelerates. These locations correspond roughly to the hotspots seen in many of the FRII objects (note again that emission from the jet itself is not included in these maps). 

At later time, after 10 Myr when the jets are turned off, the shape of the radio lobes becomes roughly spherical, resembling bubbles/cavities seen in cool core clusters. As the plasma ages, the synchrotron intensity drops due to radiative cooling and adiabatic expansion.

\begin{figure}
\includegraphics[width=\columnwidth]{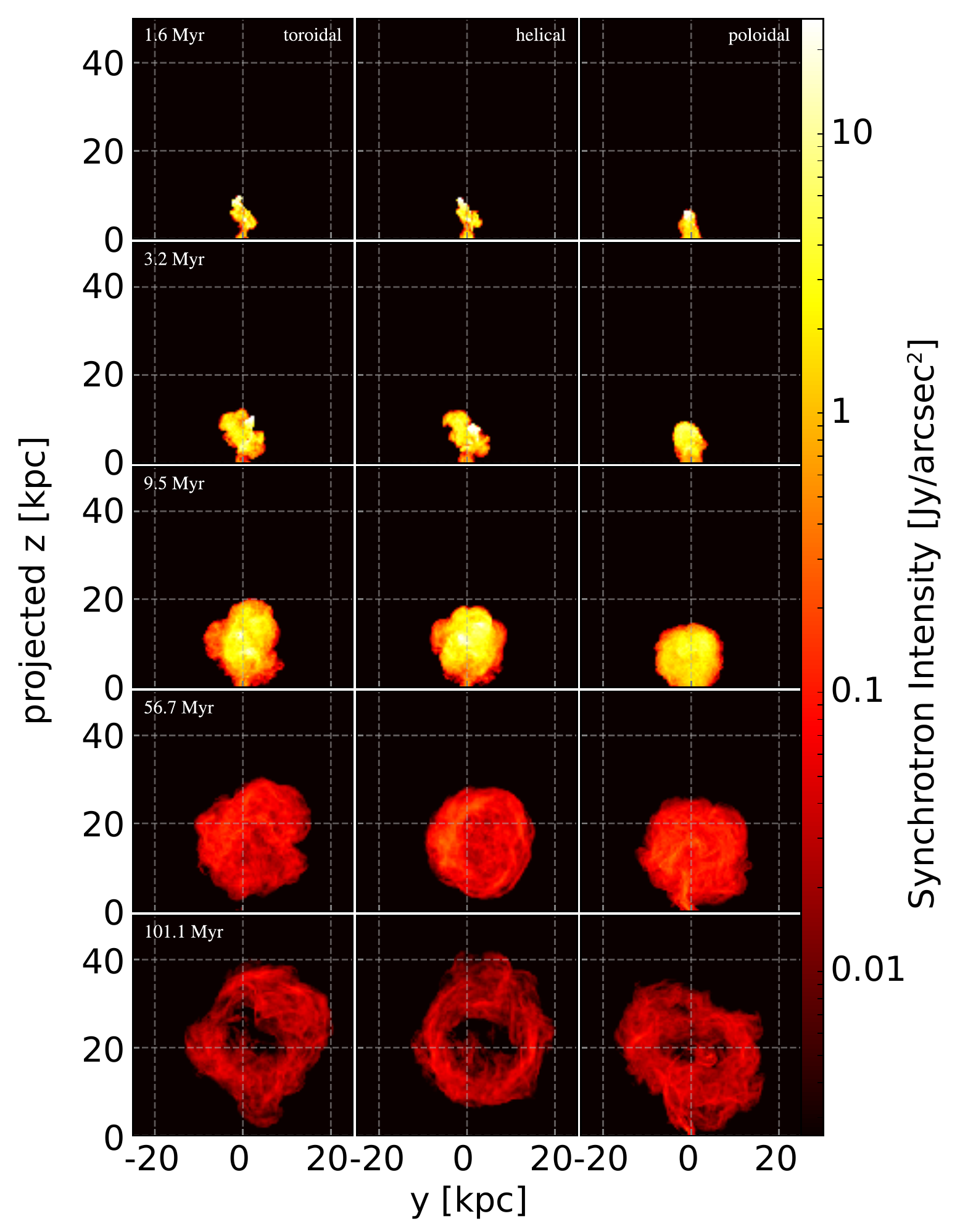}
\caption[Time series of the synthesized synchrotron maps at different times viewed from a 30-degree inclination]
	{Same as Fig.~\ref{fig:sync_150MHz} but viewed from a 30-degree
	inclination angle relative to the jet axis. Notice that the radio emitting
	plasma has a torus shape in all 3 cases that are obvious in an inclined
	angle.}
\label{fig:sync_150MHz_1_0_2}
\end{figure}

At even later times, these inflated bubbles rise further and form vortex rings, as has been discussed extensively in the literature on radio sources in cool core clusters \citep[see e.g.][]{churazov:01,gardini:07,werner:10,churazov:13}. If viewed from an inclined line-of-sight (Fig.~\ref{fig:sync_150MHz_1_0_2}), a synchrotron emitting torus is visible at 100 Myr in each simulation. As proposed by \cite{churazov:01}, the torus is formed due to the low-density hot gas rising in the stratified cluster atmosphere, subject to Rayleigh-Taylor instability.

The velocity field in such a vortex ring reflects large scale, organized rotation about the axis of the torus. The shape of the radio torus resembles, for example, the morphology of Fornax A \citep{fomalont:89,anderson:18}, which has passively evolved for at least 100 Myr and does not show any hotspots, and is also reminiscent of the "radio ear" seen in the 90cm image of Virgo A \citep{owen:00}. This structure is common among all three field-congurations, though less pronounced in the poloidal case.

\begin{figure}
\includegraphics[width=\columnwidth]{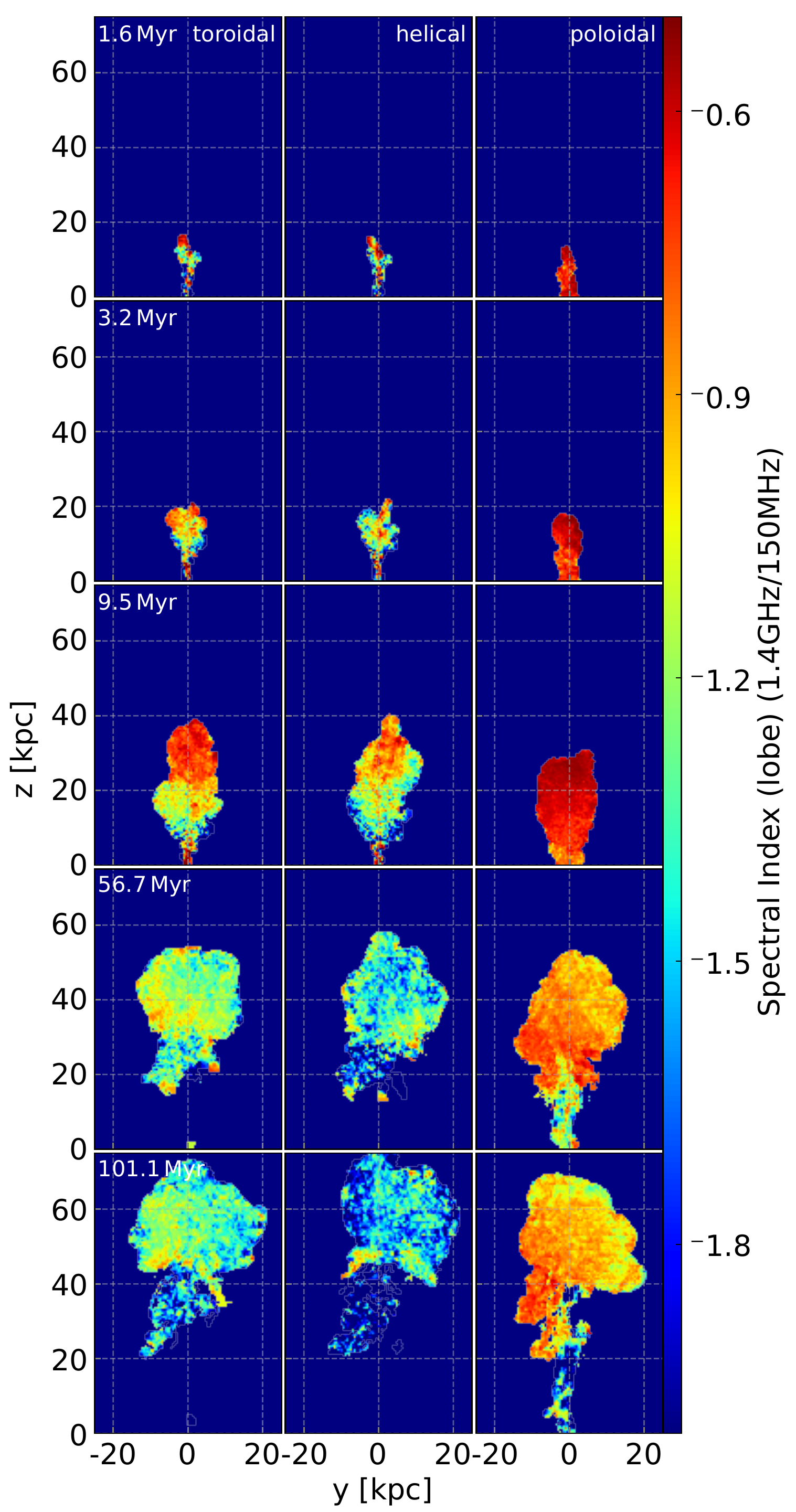}
\caption[Evolution of synchrotron spectral index at different times]
	{Evolution of synchrotron spectral index at different times in
	the simulations. The spectral index $\alpha$ is calculated as
	$\alpha = \log (I_{\nu_1}/I_{\nu_2}) / \log (\nu_1/\nu_2)$. In
	this figure, $\nu_1$ = 1400 MHz and $\nu_2$ = 150 MHz. \emph{Left} column:
	toroidal magnetic fields; \emph{middle} column; helical magnetic fields;
	\emph{right} column: poloidal column.}
\label{fig:sync_spectralindex}
\end{figure}

\begin{figure}
\includegraphics[width=\columnwidth]{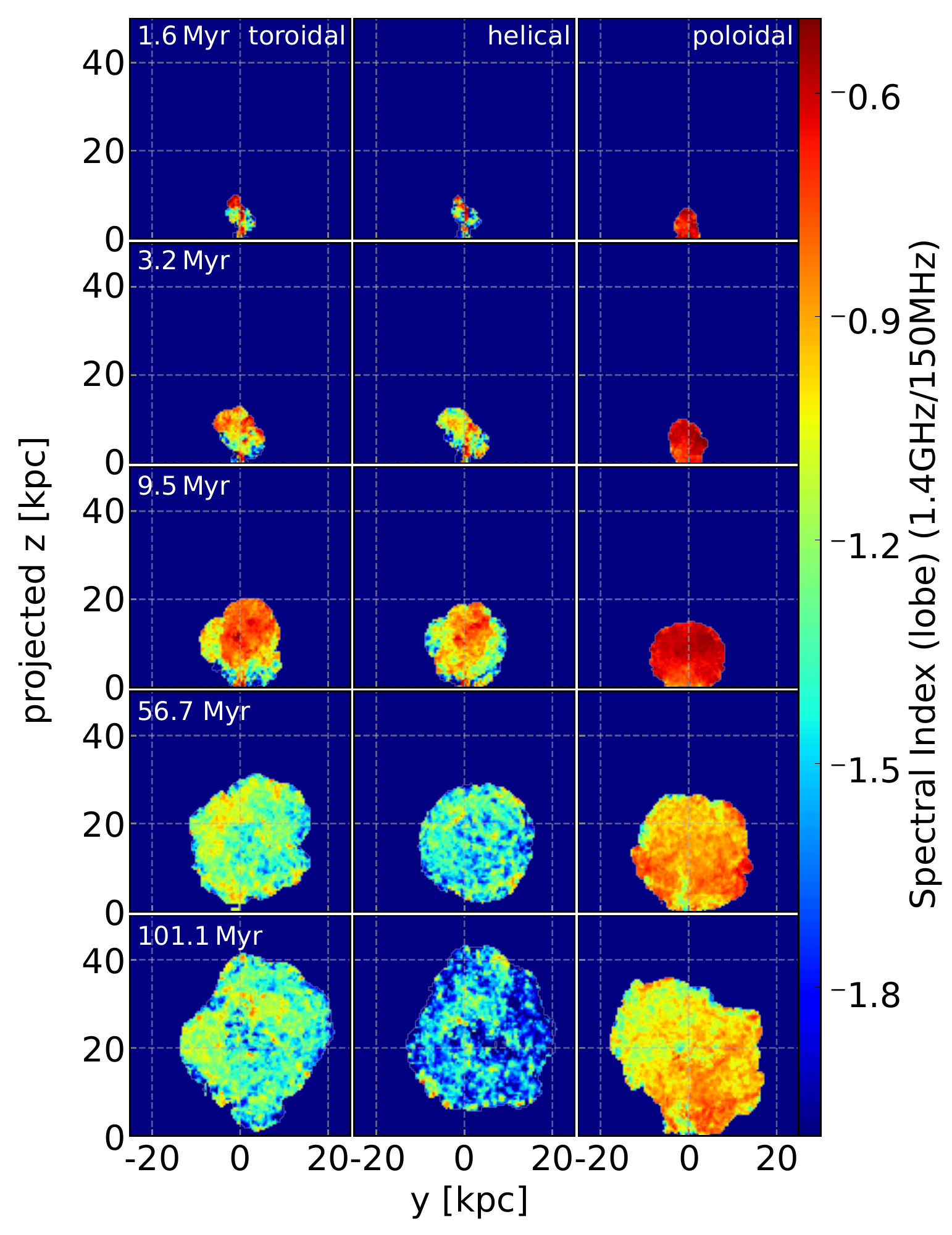}
\caption[Evolution of synchrotron spectral index at different times---inclined case]
	{Same as Fig.~\ref{fig:sync_spectralindex} but viewed from
	30-degree inclination relative to the jet axis}
\label{fig:sync_spectralindex_1_0_2}
\end{figure}

To derive a measure of spectral aging, we calculate the projected intensity at two frequencies based on the \emph{particle} aging model described in Section \ref{subsec:tracer_particles} and \ref{subsec:synchrotron_emis}. The spectral index between two frequencies is then $\alpha = \log (I_{\nu_1}/I_{\nu_2}) / \log (\nu_1/\nu_2)$. The results are shown in Figs.~\ref{fig:sync_spectralindex} and \ref{fig:sync_spectralindex_1_0_2}. A gradient from the flatter spectra at the heads to the steeper spectra at the tails is apparent in the figure when the jet is still active ($<$ 10Myr).

The synchrotron cooling tracer particle implementation adopted for these
figures assumes that the electron distribution is a fresh power-law at the working surface/termination shock, where the bulk velocity of the plasma drops and the bulk of the jet kinetic energy is dissipated in a strong shock, motivating our assumption that the synchrotron emitting plasma is accelerated in these termination shocks. This results in a young and relatively flatter spectrum ($\alpha \sim 0.5$) at the hotspots. Thus, when the jets are still active, the tips of the lobes generally show flatter spectra compared to the lower regions closer to the AGN sources. Note again that we explicitly filter out the emission from the jets. In actual observations, the removal of the active jet component can be crucial to revealing this gradient, especially in FR-I sources, where the inner lobe emission is dominated by the jet.

After leaving the hotspots, the plasma starts to cool, creating a steeper spectrum toward the inner parts of the lobes. 

Interestingly, after the jets turn off, as the radio lobe evolves, we find that the initial outside-in age gradient is not just diluted by the internal motion of the lobe plasma, but in fact reversed. This is due to the vortex formed during the rising of the bubble. The rotation of the vortex transports the young plasma down to the outside face of the vortex, while old plasma rises faster in the center due to the overall process of vortex formation. This is most apparent in the lowest two rows in Figs.~\ref{fig:sync_spectralindex} and \ref{fig:sync_spectralindex_1_0_2}. This reversal of the synchrotron age gradient could be useful when searching for and analyzing fossil radio plasma in X-ray cavities of galaxy clusters. We will discuss this in more detail in \S\ref{dis:age_gradient}.

The radio lobes in the poloidal case, compared to helical and toroidal cases, have relatively flatter spectral indices among the 3 magnetized cases. This can be seen in Fig.~\ref{fig:sync_spectralindex}, where the poloidal case has generally redder false color at every stage. This effect is most likely caused by the overall weaker magnetic field in the poloidal field case, which can be seen in Fig.~\ref{fig:mag_energy}. The cooling rates in the poloidal case are thus relatively slower, yielding flatter spectra when compared to the other 2 cases at the same time. On the other hand, the helical case has the steepest spectral index because of the stronger magnetic fields. Note that although we inject more magnetic energy in the toroidal case, it does not have the strongest magnetic fields because of the onset of the kink instability and the resulting numerical reconnection discussed in Section \ref{sec:kink_instability}.

%\begin{figure}
%\includegraphics[width=0.7\columnwidth]{synchrotron_spectra.pdf}
%\caption{Total synchrotron spectra evolution over time.}
%\label{fig:sync_spectra}
%\end{figure}

\section{Discussion}
\label{sec:discussion}

\subsection{Magnetic Field Strength and the Development of Kink Instability within the Jet}
\label{sec:kink_instability}

We find that the helical jet with jitter overtakes the purely toroidal geometry to become the fastest propagating jet head at later times. As we show below, the helical case also maintains the highest total magnetic energy among all three magnetized cases, while the poloidal case has the lowest. Finally, we saw that the hydrodynamic case slows down to speeds comparable to the poloidal case despite the lack of the additional lateral magnetic pressure. We understand these outcomes in the context of different dynamical instabilities the simulated jets are subject to, and the resulting differences both in jet propagation (which is a physical effect) and numerical dissipation (which is, at root, an un-physical effect that is unavoidable in any grid-based simulation, but sometimes similar to turbulent dissipation encountered in astrophysical systems).

Magnetic energy injection occurs differently for the different cases: For the poloidal case, the initial dipole field is stretched and amplified and no additional magnetic flux is injected in the simulation domain. On the other hand, we inject additional toroidal magnetic flux into the domain for the toroidal and helical cases that is constantly transported out of the nozzle.

Due to the lateral expansion of the jets, the toroidal ($\phi$) component scales as $1/r$ while the poloidal ($z$) component scales as $1/r^2$. For initially comparable fields in toroidal and poloidal components, once the jet expands, the toroidal component will dominate the overall magnetic field {\em in the jet}.

\begin{figure}
\includegraphics[width=\columnwidth]{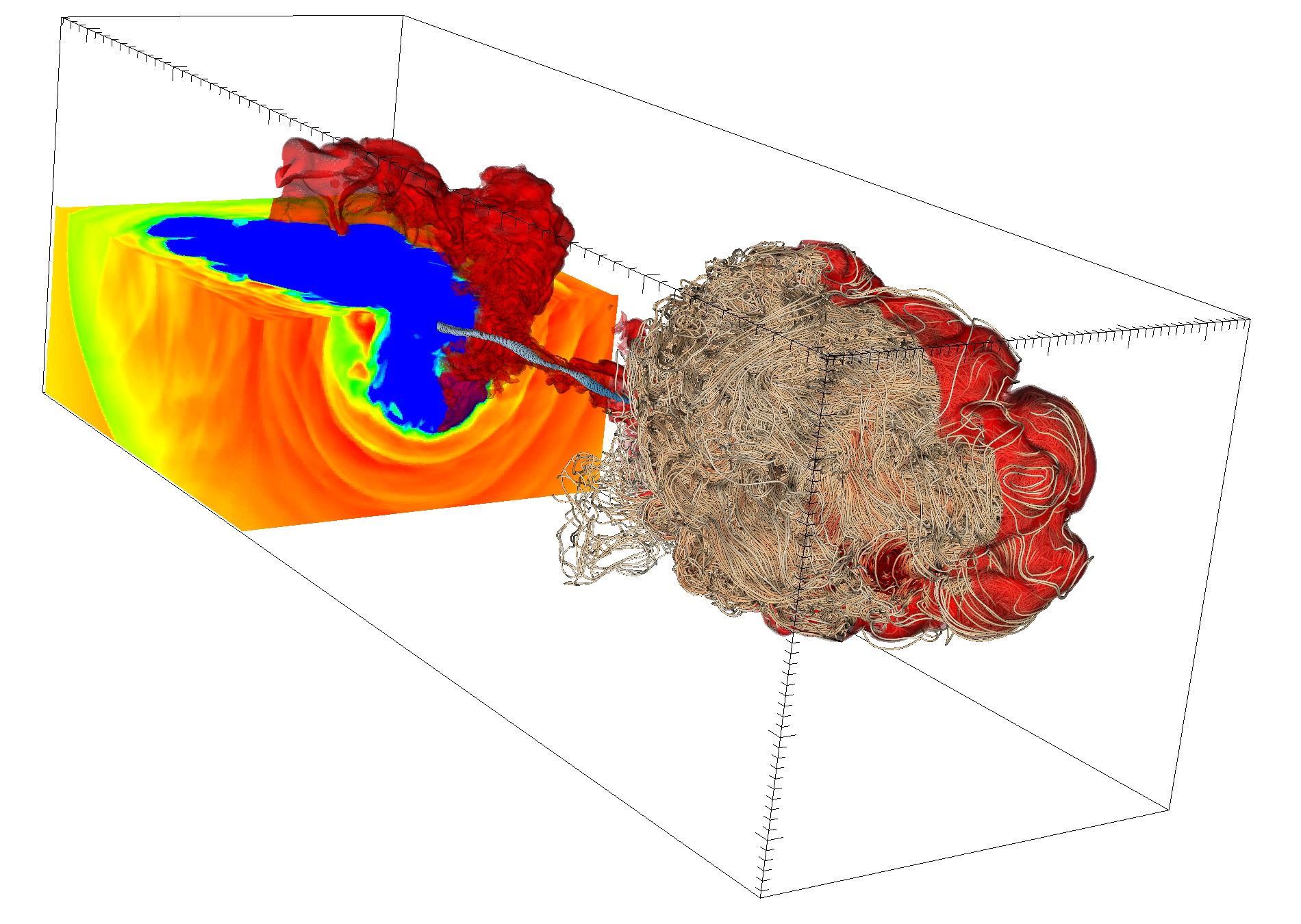}
\caption[3D rendering of a simulation snapshot with helical magnetic fields]{3D rendering of a simulation snapshot with helical magnetic fields in the jets. At left, slices show the temperature map (hot in blue and cold in red). At right, streamlines of magnetic fields represent the magnetic structure inside the lobe. Although the magnetic fields are highly ordered in the jet, they become mostly tangled in the lobe.}
\label{fig:h1_rendering}
\end{figure}

\begin{figure}
\includegraphics[width=0.99\columnwidth]{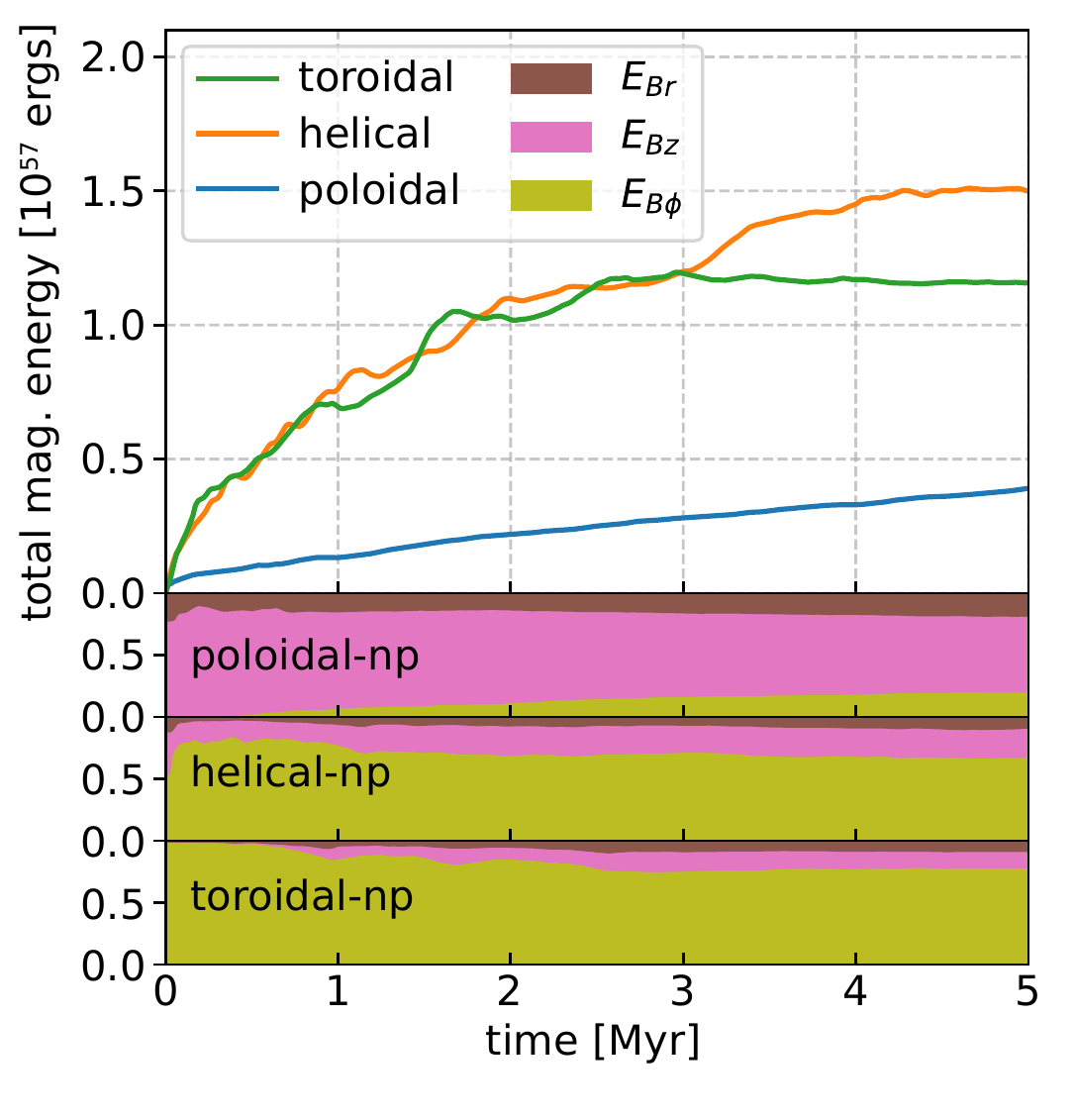}
\includegraphics[width=0.99\columnwidth]{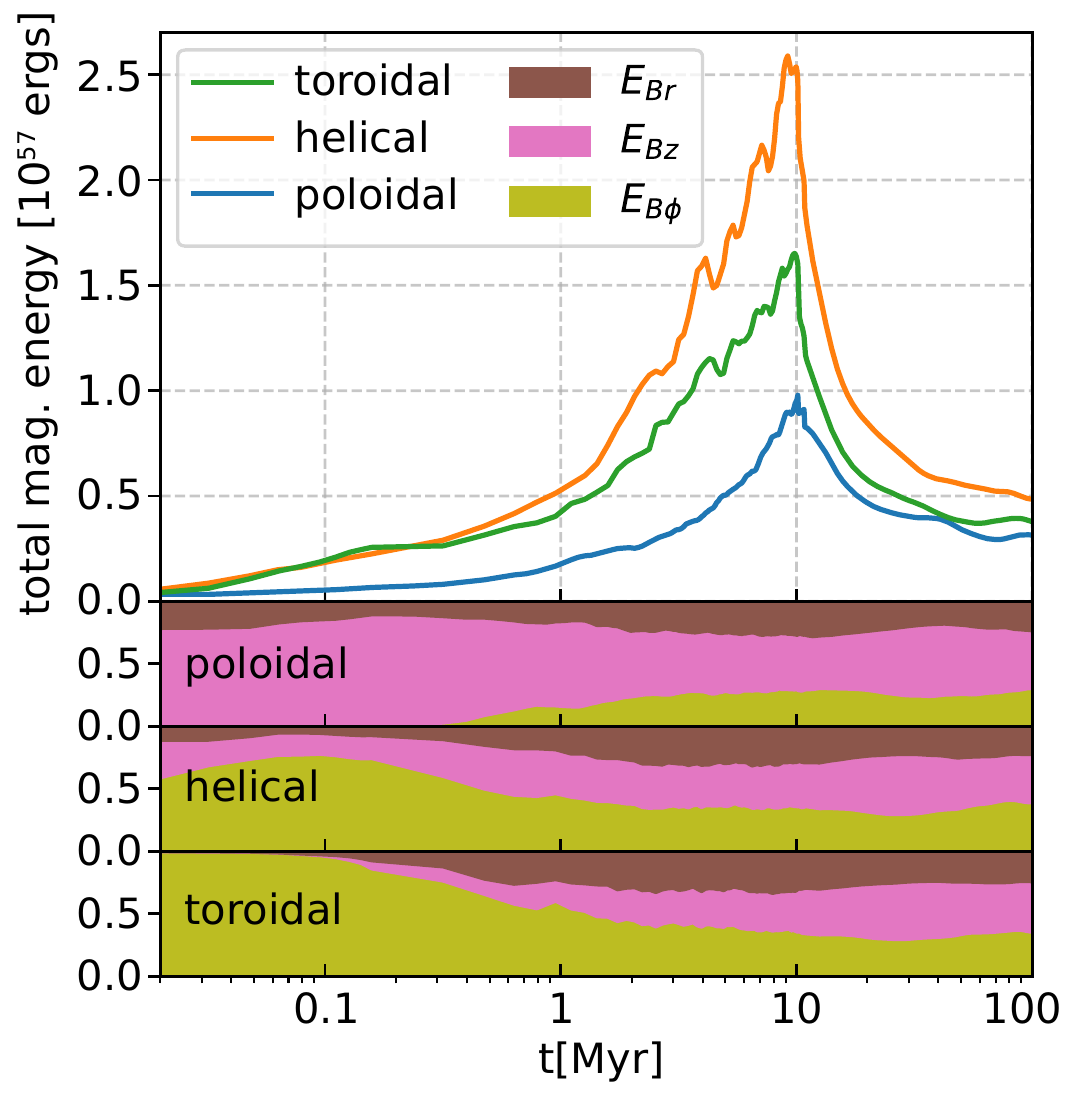}
\caption[Evolution of the total magnetic energy and different components]
	{Evolution of the magnetic energy and different components of the magnetic energy in the simulations. Different colors denote different magnetic configurations of the injection nozzle of each simulation. Solid lines indicate the total magnetic energy, while the filled charts show the percentage of the $r$, $z$, and $\phi$ components. \emph{Top panel}: non-precessing simulations (denoted ``np``). Although in toroidal case more magnetic flux is injected in the domain, at later times the amount of magnetic energy is surpassed by the helical case. The poloidal case remains the lowest in magnetic energy. \emph{Bottom panel}: precessing jets active for 10 Myr. The magnetic energy decreases after the jet is off due to (mostly numerical) reconnection in the lobes. Note that at later times (after 10Myr), the $z$ component increases, which is likely caused by the buoyantly rising bubble stretching the magnetic fields vertically.}
\label{fig:mag_energy}
\end{figure}

{\em In the lobes}, magnetic fields become mostly tangled due to the turbulent motions. This can be seen in the rendered image in Fig.~\ref{fig:h1_rendering} as well as the evolution of different magnetic components in Fig.~\ref{fig:mag_energy}. In order to investigate the evolution of the magnetic field, we measure the magnetic energy of the $(r, z, \phi)$ components in cylindrical coordinates, with the $z$-axis parallel to the mean jet axis. In the non-precessing cases (denoted ``np``), the $\phi$ component is maintained at a high level for both toroidal and helical cases, while in the precessing cases, the $\phi$ component is no longer dominant energetically. Rather, the different components are comparable to each other, especially around and after 10 Myr when the jet is turned off as shown in the bottom panel of Fig.~\ref{fig:mag_energy}. This is likely due to the turbulent nature of the flow within the radio lobes, which efficiently mixes the different modes.

\begin{figure}
\includegraphics[width=\columnwidth]{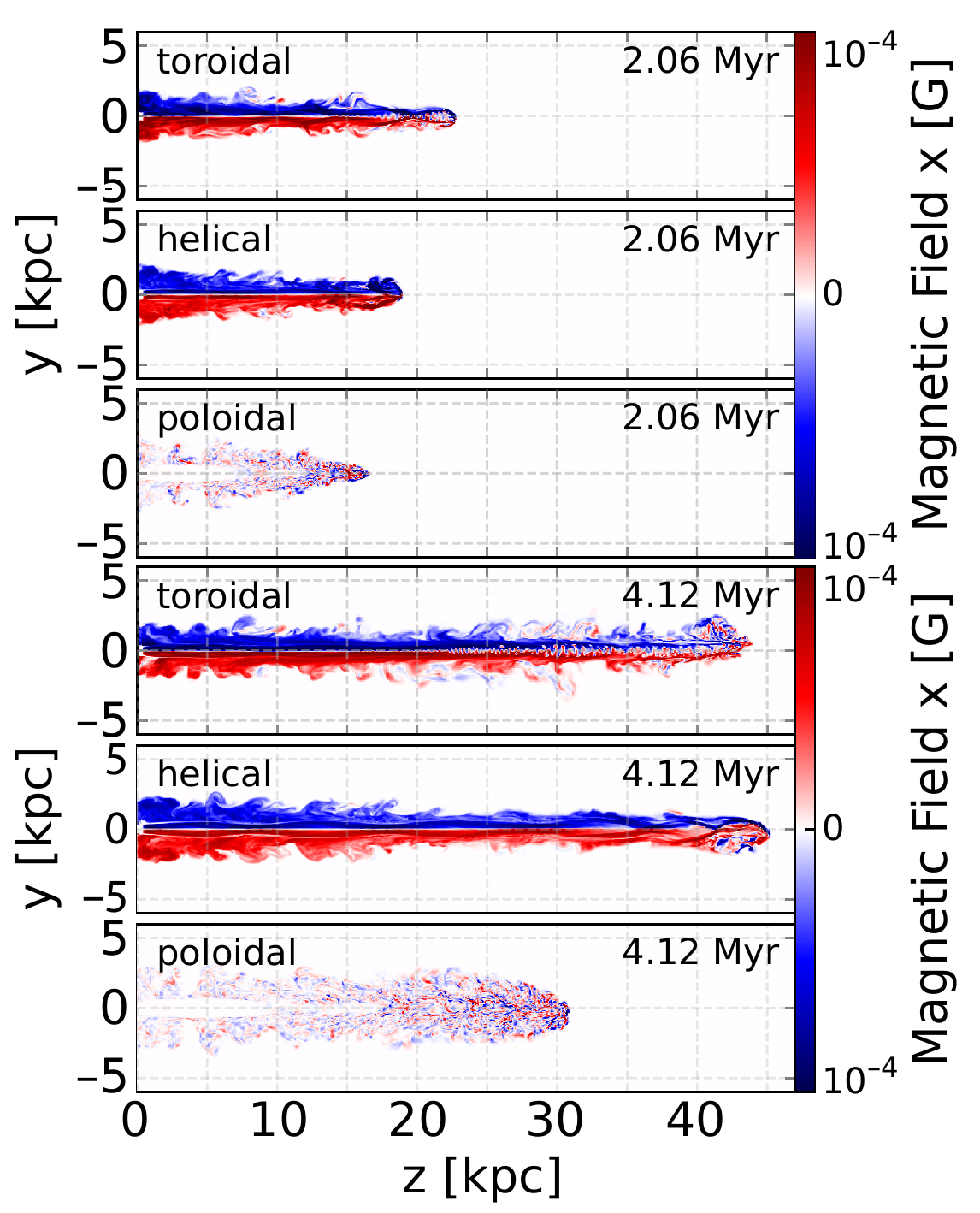}
\caption[Comparison for slices of magnetic fields]
	{Central slices of the magnetic fields in the x-direction (perpendicular to the image plane). In the toroidal case, kink instabilities start to develop {{beyond 15 and 20 kpc at 2.06 and 4.12 Myr (top and bottom three panels, respectively)}}. Due to the limited resolution in our simulations, the instabilities result in magnetic energy being converted to thermal energy.	Thus, it reduces the jet head propagation velocity of the toroidal case at
	later times.}
\label{fig:compare_3_nojiggle_Bx}
\end{figure}

The long-term behavior of the helical and the purely toroidal configuration require further discussion. We find that the counter-intuitive behavior described above can be explained by the action of the kink instability, which can also account for the field decrease in the toroidal case. This can be clearly seen in Fig.~\ref{fig:compare_3_nojiggle_Bx}. The toroidal ($\phi$) field in the toroidal case, although initially stronger than in the helical case, suffers from dynamical disruption, effectively widening the jet cross section, and subsequent numerical dissipation of the magnetic field loops due to the kink modes that become non-linear along the jet.

\cite{appl:00} pointed out that current-driven instabilities are important in astrophysical jets and carried out linear stability analysis of the growth rates. They found that the growth rate depends on the radius of the jet, the Alfv\'en velocity, as well as the magnetic pitch $P = r B_{z}/B_{\phi}$. When normalized by the Alfv\'en crossing time ($r_{\text{jet}}/v_A$), the growth rate is inversely proportional to the pitch or third power of the pitch depending on small or large pitch regimes. Larger pitch leads to a lower growth rate, i.e. the kink instability is suppressed. Although in our simulations the pitch profile is different from the profiles analyzed in \cite{appl:00}, we can still compare our instability growth rate to their calculations.

First, we consider the toroidal case, which should have a magnetic pitch close to zero and develop instabilities very quickly. However, we do not see the growth of the instability until the jet reaches beyond 18 kpc (Fig.~\ref{fig:compare_3_nojiggle_Bx}). This can be accounted for by the numerical resolution required to resolve the unstable modes in our simulations. Roughly speaking, the minimum wavelength that can be resolved and thus grow in the simulation is about 4 cell-widths, which is about a quarter of the jet diameter. The maximum wavenumber that can be resolved is then $k_{\text{max}} = 2 \pi / \lambda_{\text{min}} \sim 12 / r_{\text{jet}}$. The growth time of this unstable wavenumber is about $ 3 r_{\text{jet}} / v_A$ (c.f. Fig. 3 in \cite{appl:00}) or $\sim 0.2$ Myr given $ r_{\text{jet}} \sim 240$ pc and $v_A = 0.012c$ in our simulations. The jet fluid is moving at 0.1 $c$ in our simulations, which gives an instability e-folding length scale of 6 kpc. The instability becomes visible at about 3 times this length scale.

Second, we consider the helical case, in which the jet has magnetic pitch of order of 1 ($P/r_{\text{jet}} \simeq 1$). From Fig. 3 in \cite{appl:00}, we can infer that the maximum growth rate is about 0.1 ($\Gamma r_{\text{jet}}/v_A \sim 0.1$) and thus the e-folding growth time is of order 10 Alfv\'en crossing times ($ t = 1/\Gamma = 10 \, r_{\text{jet}} / v_A$), which is roughly 0.8 Myr in our cases. Considering the jet fluid is 0.1 $c$, this timescale corresponds to a length scale of 24 kpc. We would expect the instability to become apparent at 3 or 4 times of this length scale. This can explain why there is no visible instability for the helical jet in Fig.~\ref{fig:compare_3_nojiggle_Bx}.

This is similar to the findings of previous works \citep{nakamura:07,guan:14,bromberg:16,barniol-duran:17} that if the ratio between toroidal and poloidal fields is large, internal kink modes will develop and disrupt the jets. Due to the limit of our numerical resolution, the kink instability drives numerical reconnection at the core of the jet, which suppresses the magnetic field strength in the toroidal case below that of the helical case.

In \cite{bromberg:16,barniol-duran:17}, the authors attribute the development of the kink instability to the external medium density profile since the pitch angle of their jets depends strongly on the external medium. They argue that the jets propagating in denser external medium (headed jets) suffer from the kink instability, while the jets propagating in the less dense medium or a previously evacuated funnel (headless jets) can maintain their stability. Various factors, e.g. the spin of the central black hole, can change the magnetic pitch. In this work, we confirm that even in the same density profile, different magnetic pitches can still result in varying growth rates of the kink instability.

Finally, in the hydrodynamic case, in which magnetic fields are absent, we see the development of Kelvin-Helmholtz (KH) modes at the boundaries of the cavity. The magnetized jets and lobes suppress the KH instability at the shearing boundaries. This is consistent with previous findings that magnetic fields can suppress the instabilities \citep[see e.g.][]{ryu:00,jones:05}.

\subsection{Comparison to Cygnus A}

Given the obvious visual similarities, it is instructive to compare our synthesized radio maps with Cygnus A, the brightest and nearest powerful FR II radio source \citep{carilli:96}. Although our simulations are not specifically modeled to resemble the Cygnus A system, the resulting radio maps have a lot of similarities to the observed properties of the proto-typical FRII object, including the morphology as well as spectral index distribution \citep{mckean:16}.

In Fig.~\ref{fig:sync_150MHz}, single or multiple hotspots are visible when the jet is active (before 10 Myr). The misalignments between jet-axis and hotspots are natural results of the jitter implementation and mimic the two hotspots seen in Cygnus A \citep{pyrzas:15}. While active jets are seen in Cyg A, we compare only the lobes with our synthesized image since we leave out most of the jet emission. At 50 Myr, more filamentary structures are developed, similar to Cygnus A. We note that the estimated age of the Cygnus A ranges from few Myr \citep{carilli:91} to 100 Myr due to the large uncertainties in the magnetic fields. Differences in structure could also arise from the different duty cycles of the source.

Our spectral index maps at 10 Myr show a strong gradient from the hotspot to the tail of the lobe. The head of the lobe, where the plasma is freshly injected, has a flat spectrum with spectral index $\alpha$ close to 0.5, while the spectral index steepens toward the tail with $\alpha \sim 2$, where most of the plasma is old. This is consistent with the detailed observations of Cygnus A \citep{carilli:91}.

Clearly, the morphology of the jets simulated here differs from the typical FRI-type morphology of cool-core cluster radio sources like Perseus A and Virgo A. It worth emphasizing that, while the true nature of the difference between FRI and FRII sources is still unsettled, it is likely rooted in physics we did not include in our simulations (e.g., proper special relativistic MHD with bulk Lorentz factors of order 5 or higher, mass loading by entrainment, e.g., \citealt{bicknell:96}). However, the goal of our simulations is not a full reconstruction of a particular system, but rather a broader investigation of the impact of different field topologies on the properties of radio galaxies.

\subsection{Spectral index gradient evolution}
\label{dis:age_gradient}

In Section \ref{subsec:synchrotron_properties}, we see a clear spectral index gradient while the jets are still active when the sources are observed close to perpendicular to the jet axis. During the active jet phase (shown at the time of 9.5 Myr in the relevant figures), we see a flatter spectral index/younger plasma located at the tips of the lobe, while the older plasma is at the tails close to the central plane. This gradient of the spectral index has been seen in many observations of FR-II sources \citep[e.g.][]{steenbrugge:10,shulevski:17,savini:18}. While it is more difficult to establish such global trends in FR-I sources, given the subtraction of the jet emission from the diffuse emission from the inner radio lobe, it has been observed in some FR-I sources \citep[e.g.][]{kolokythas:15}.

At a later time after the jet turns off (roughly at 50 Myr as shown in the figures), we find an intriguing reversal of the age gradient in the spectral index map. The relatively younger plasma is now at smaller radii of the lobe, while the outermost parts of the lobes appear to be older, as shown in Fig.~\ref{fig:sync_spectralindex}.

If the viewing direction is closer to the jet axis, the spectral index is flatter in the center and gets steeper closer to the edges of the lobe while the jet is active, as shown in the third row (9.5 Myr) of Fig.~\ref{fig:sync_spectralindex_1_0_2}. However, once the jets shut off, the distribution of the spectral index becomes again reversed -- steeper at the center and relatively flatter closer to the edges, as in the 4th row (56.7 Myr) of Fig.~\ref{fig:sync_spectralindex_1_0_2}.

This reversal can be understood as a consequence of rotation within the rising bubble. Our radio sources resemble typical FR-II sources, in which the radio emission is brighter away from the central engine. The fresh, younger plasma appears in the vicinity of the hotspots which are located at the farthest ends of the radio lobes. While it is still not clear what acceleration mechanism powers FR-II sources --- whether the relativistic electrons are accelerated in the terminal shocks, i.e. the hotspots, or \textit{in situ} in the jets ---, those sources produce a spectral index (or age) gradient from flatter (younger) at the farther ends of the lobes to steeper (older) closer to the central source.

\begin{figure}
    \centering
    \includegraphics[width=0.98\columnwidth]{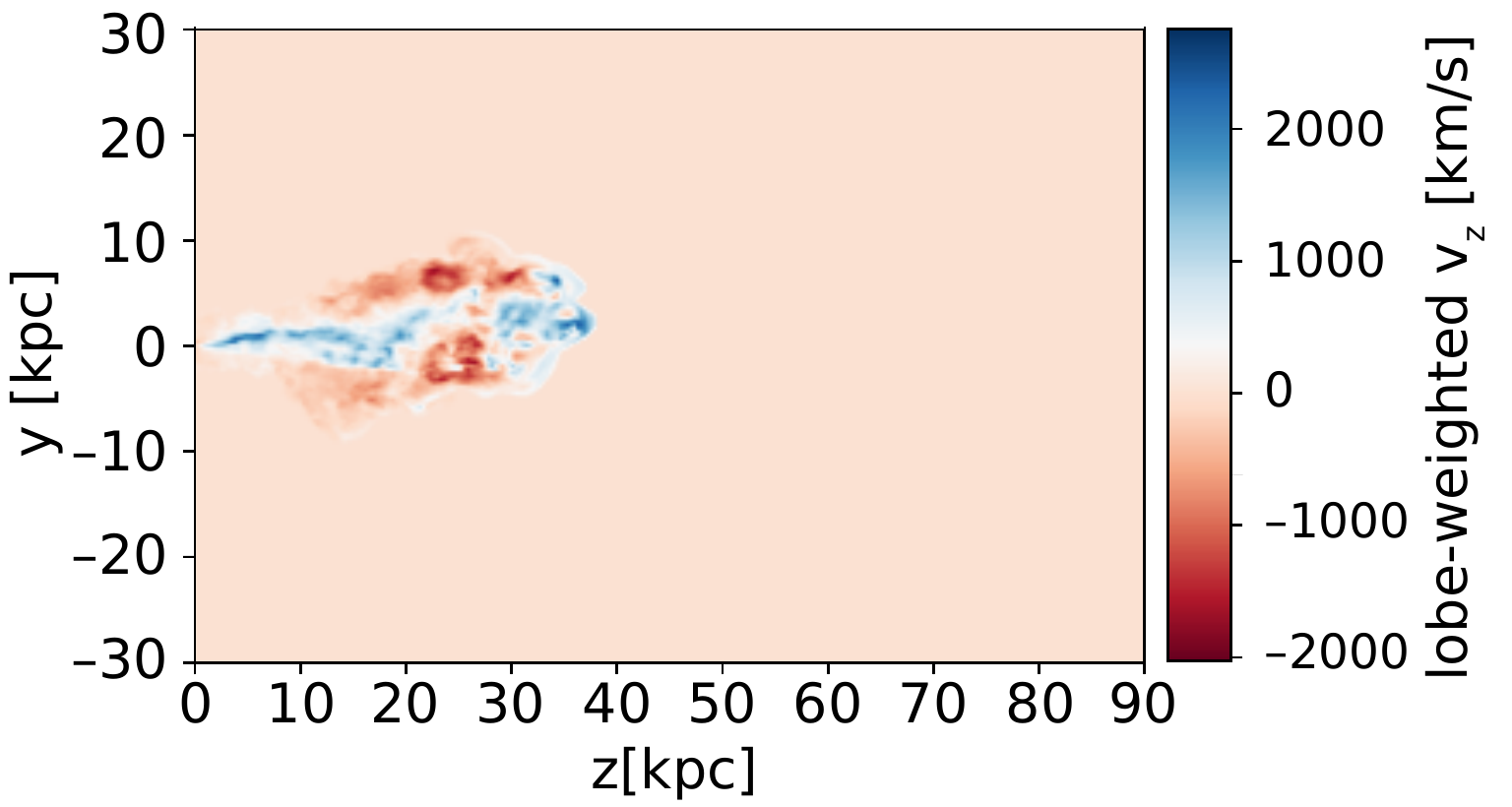}
    
    \includegraphics[width=0.98\columnwidth]{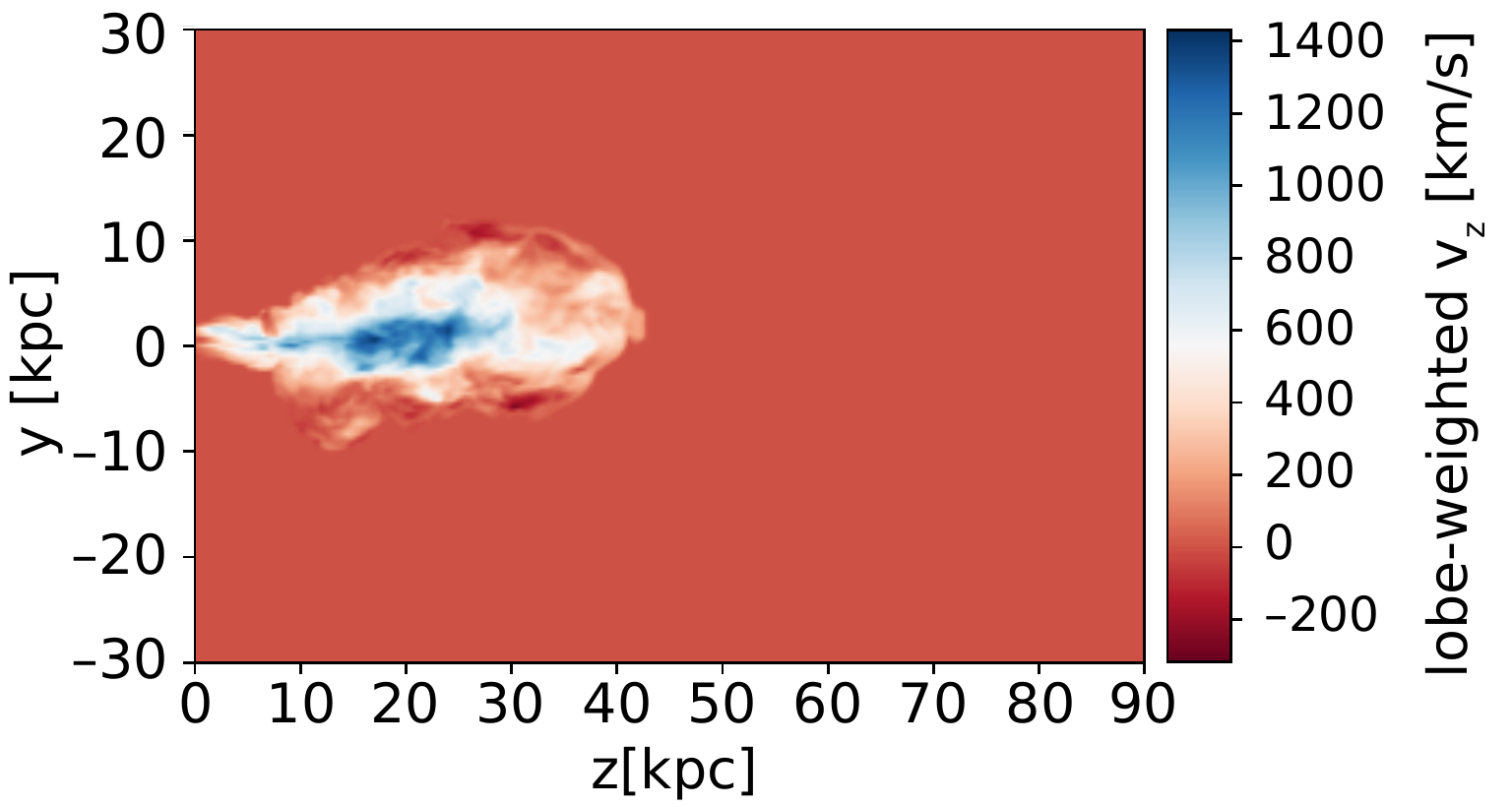}
    
    \includegraphics[width=0.98\columnwidth]{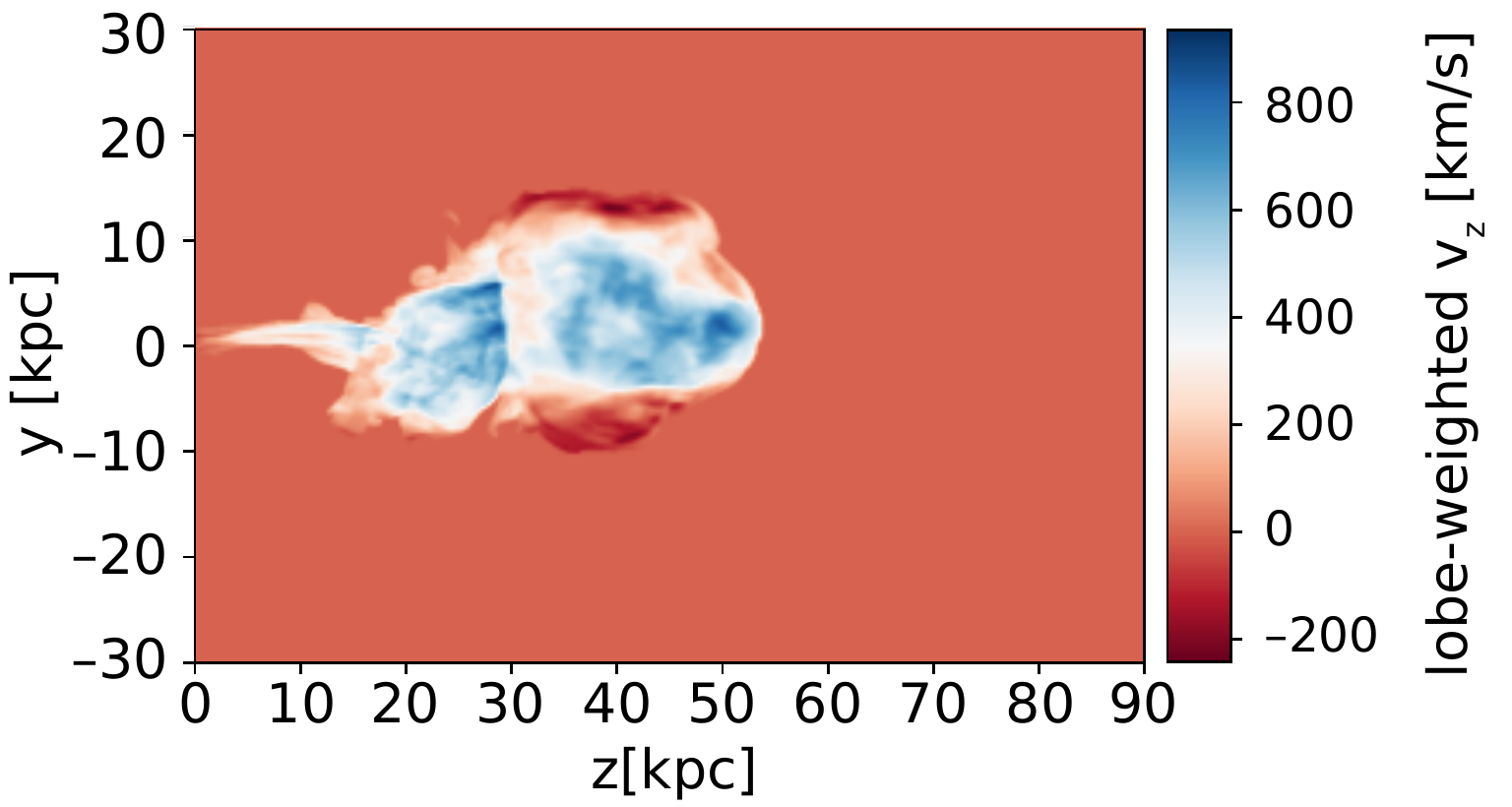}
    
    \includegraphics[width=0.98\columnwidth]{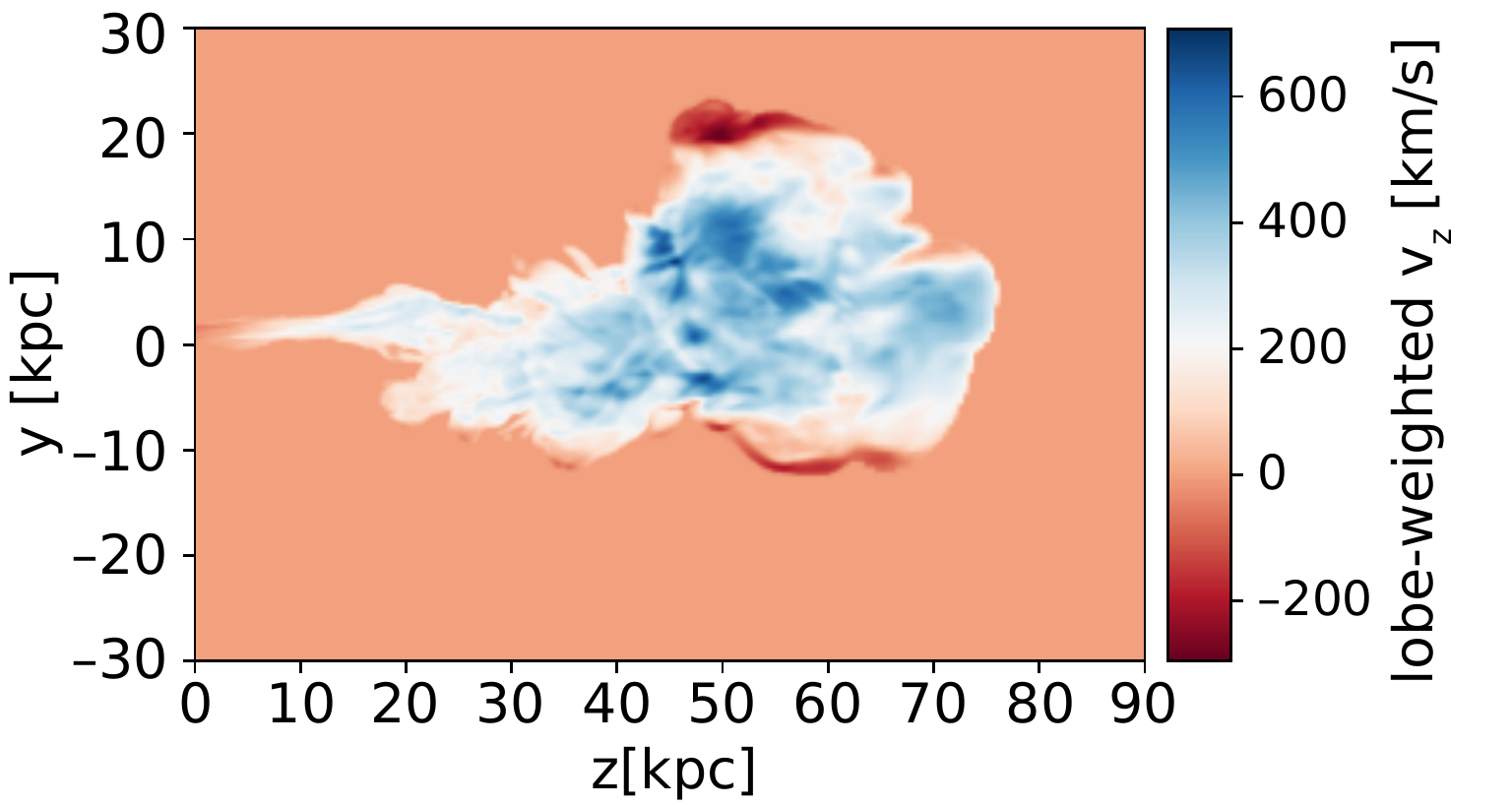}
    \caption{Large scale circulation within the radio lobes of the helical field simulation. Shown in each panel: Average z-velocity, weighted by the jet tracer fluid mass fraction for 11, 21, 51, and 101Myrs from top to bottom, each showing upwelling along the central spine of the lobe and down-flow along the sides. Initially, this backflow motion is driven by the pressure-gradient within the cocoon. At later times, the circulation is driven by buyancy. Both effects contribute to the reversal of the age gradient seen at later times in the simulation. Note that the net z-velocity of the lobe is positive; velocities that are downward in the frame of the lobe still have positive net z-velocity in the frame of the observer. {{The limits on the color bar were chosen arbitrarily to optimize the dynamic range of each figure.}}}
    \label{fig:vz_proj}
\end{figure}

\begin{figure}
    \centering
    \includegraphics[width=0.98\columnwidth]{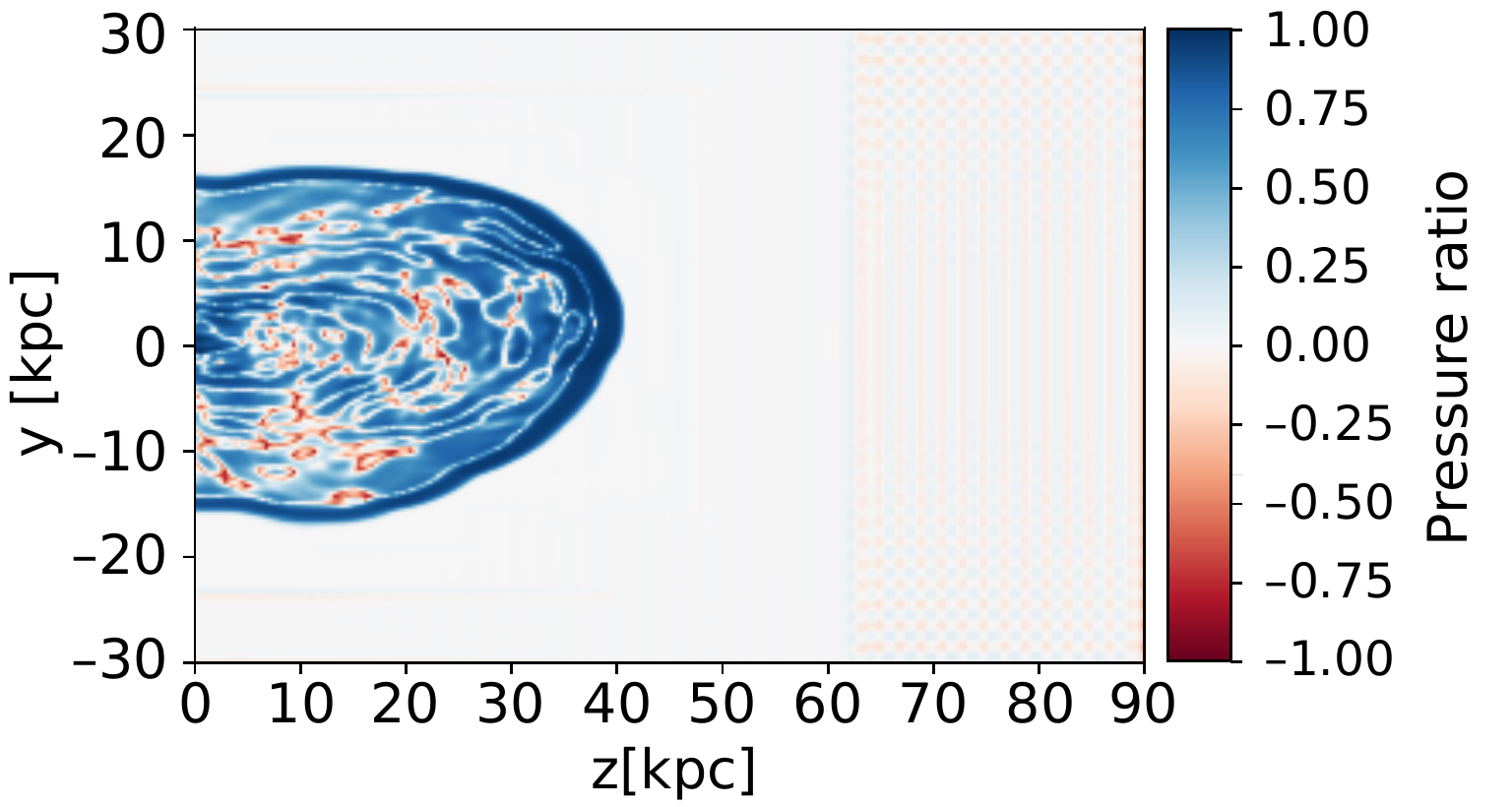}
    
    \includegraphics[width=0.98\columnwidth]{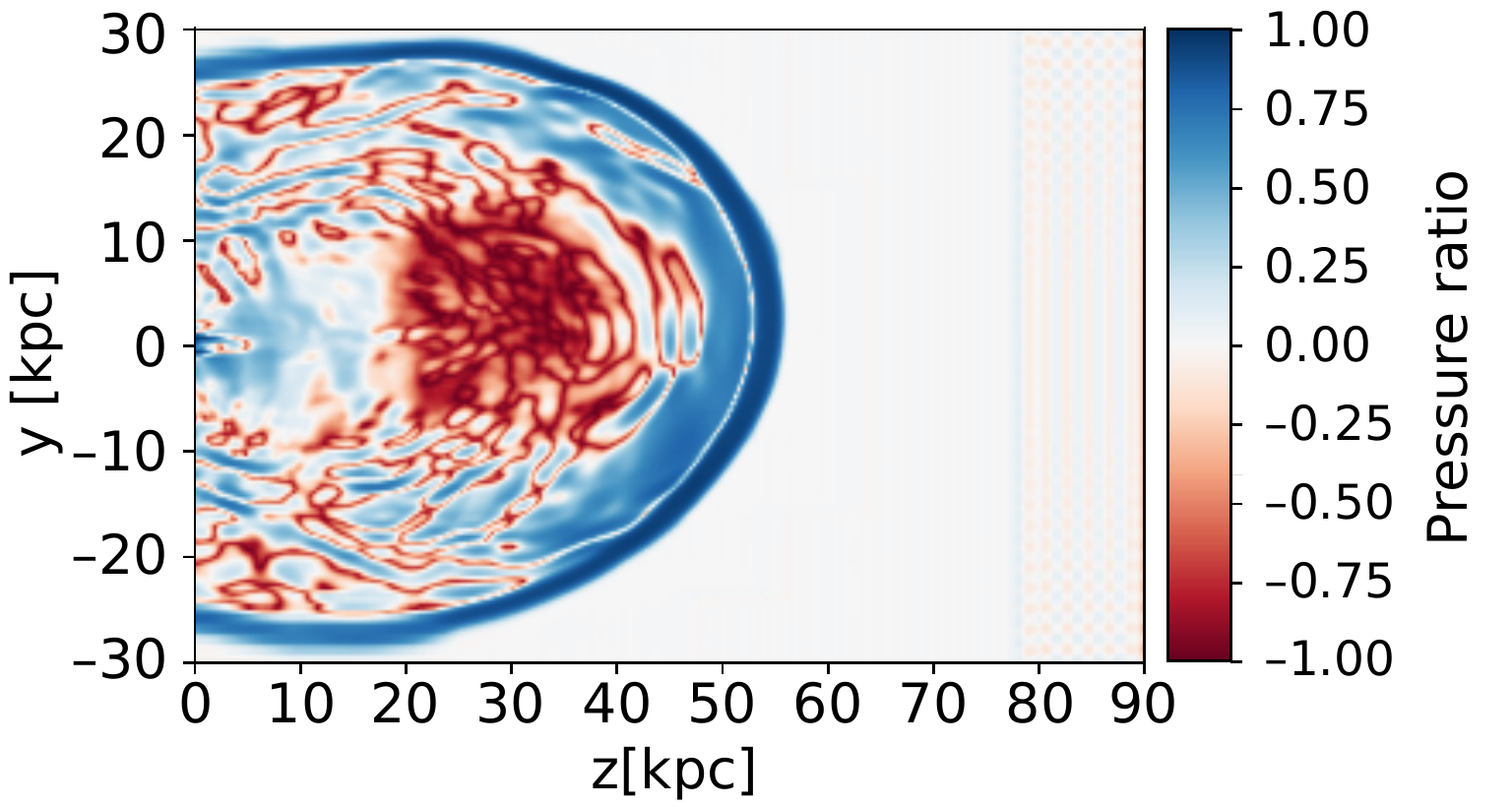}

    \includegraphics[width=0.98\columnwidth]{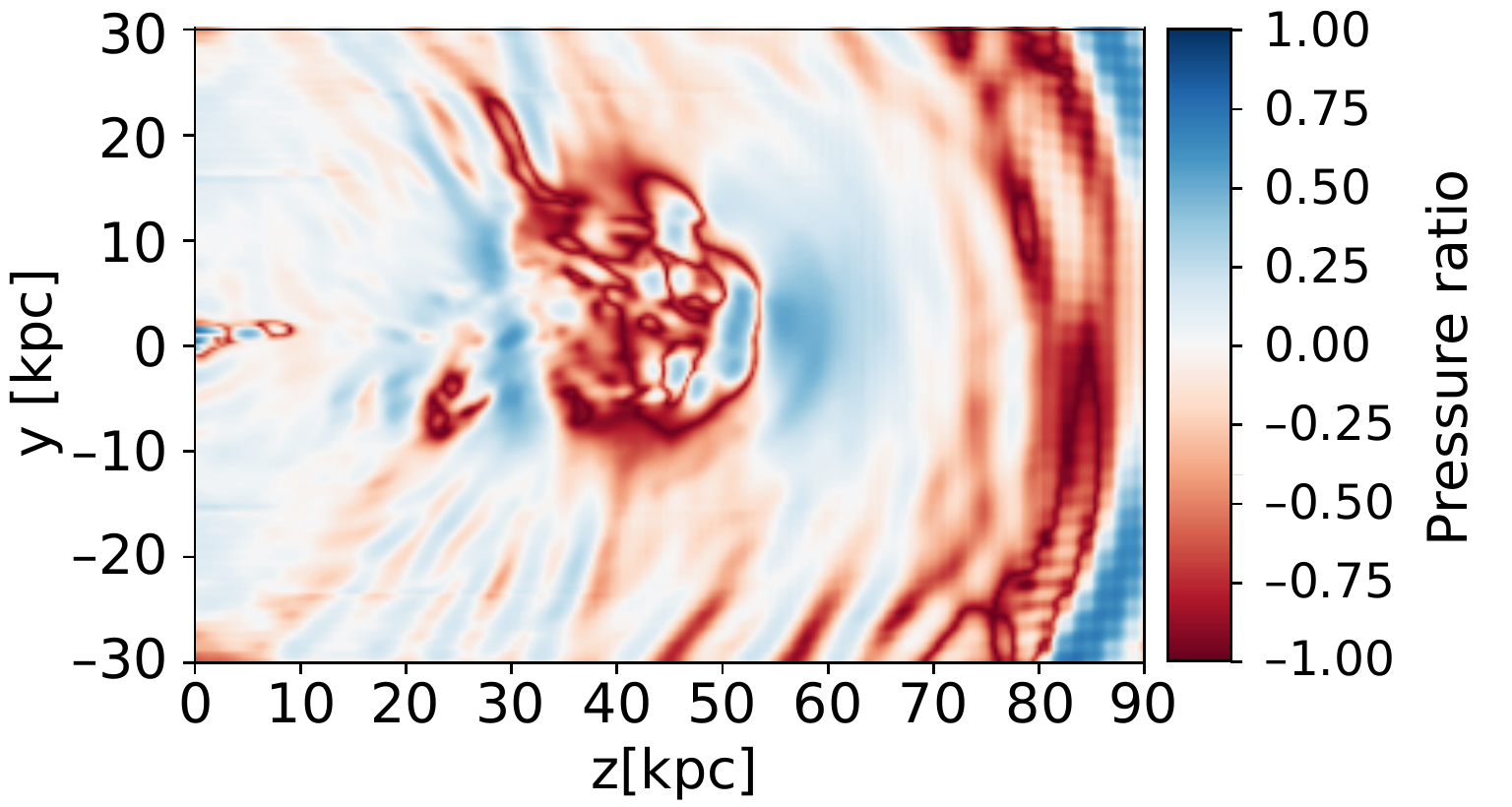}

    \includegraphics[width=0.98\columnwidth]{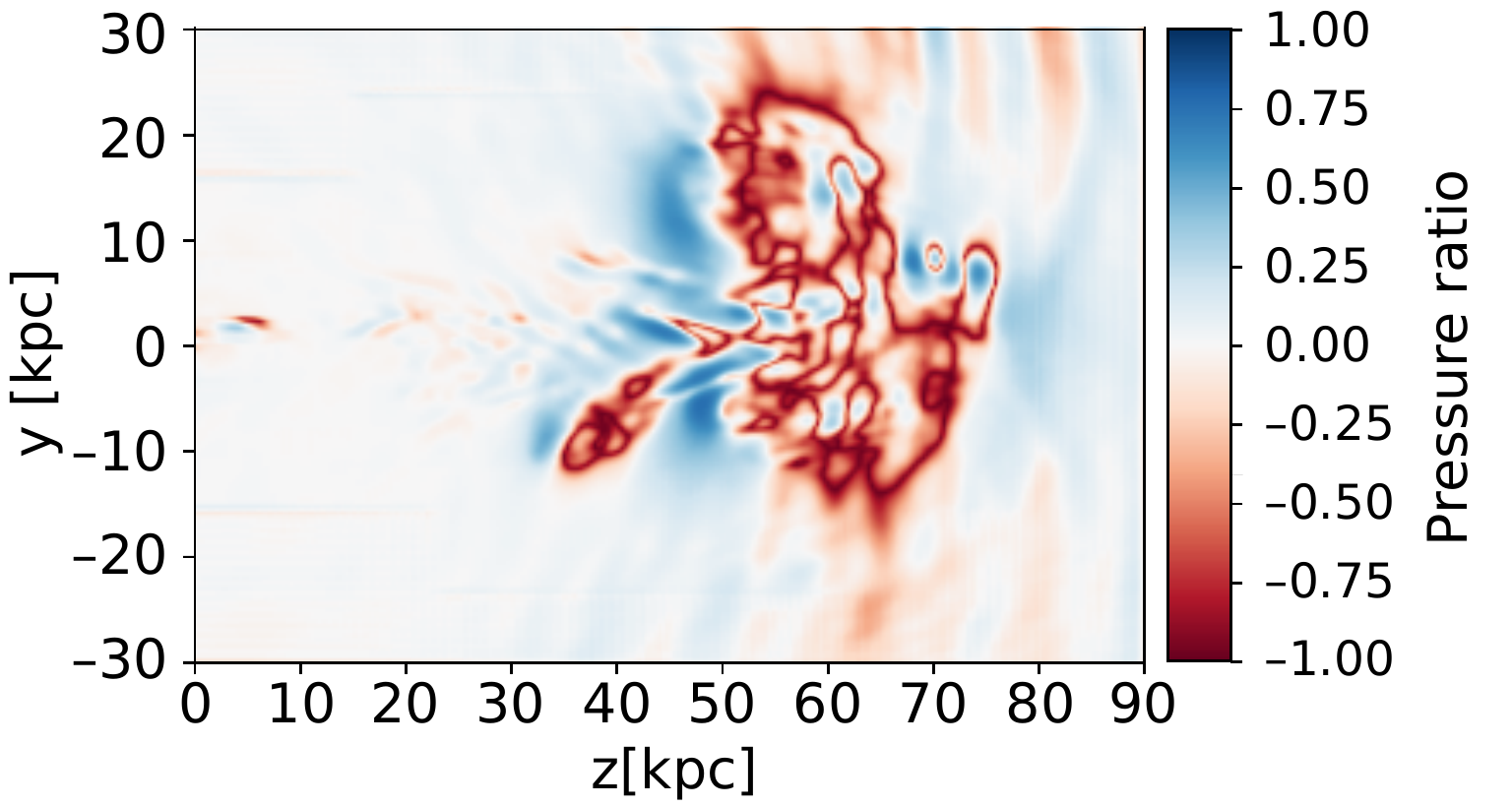}
    \caption{Slice of the net buoyancy force balance corresponding to the velocity panels in Fig.~\ref{fig:vz_proj}. The figure shows that the initial circulation is not driven by buoyancy, as the entire cocoon is strongly overpressured and expanding. However, by 50Myrs, the lobes are in approximate pressure equilibrium with the ICM on average and buoyancy is the driving mechanism for large scale motion.}
    \label{fig:pressure_ratio}
\end{figure}

Two effects contribute to this rotation:
\begin{itemize}
    \item{First, the injection of the jet creates a backflow toward the tail of the lobe, which has been pointed out and extensively discussed by numerous authors \citep[e.g.,][]{norman:82,antonuccio-delogu:10,cielo:17}. Depending on the size and location of the cavity, backflows might reach back to the equatorial plane where the supermassive black hole is located, or circulate within the cavity. In our simulations, the helical and toroidal cases have cavities away from the central plane, while the poloidal and hydro cases show more puffy cavities extending to the equatorial plane (Fig.~\ref{fig:compare_4}), due to the differences in internal collimation between these cases. However, in all cases, a backflow exists, driven by the pressure gradient within the cocoon. Figure \ref{fig:vz_proj} shows the net circulation generated within the radio lobes of the helical simulation 1 Myr after the jet terminates, at 11Myrs.} 
    \item{Second, the buoyantly rising bubbles naturally evolve into vortex rings that drive the global rotation of the plasma in the bubble about the ring  \cite[e.g.][]{churazov:01}. The sense of rotation is such that the plasma moves away from the central source (down the gravitational potential) along the axis of the vortex (i.e., the mean jet direction), and downwards along the outside of the vortex ring. That is, the rotation twists the plasma in a double-right-handed way. Because of the ongoing buoyant rise of the lobes/bubbles, the vortex z-velocity still has positive net values in most of the lobe, however, in the frame of the bubble, the gas is moving downward. Figure \ref{fig:pressure_ratio} shows the importance of buoyancy forces compared through what we term the force balance $FB$, defined as\begin{equation}
        FB\equiv\frac{\left|\nabla P\right| - \left|\rho \vec{g}\right|}{\left|\nabla P\right| + \left|\rho\vec{g}\right|}
    \end{equation}}
    Positive and negative values of $FB$ indicate regions where pressure forces or gravity (buoyancy) forces dominate the momentum equation, respectively.
\end{itemize}

\begin{figure}
\includegraphics[width=\columnwidth]{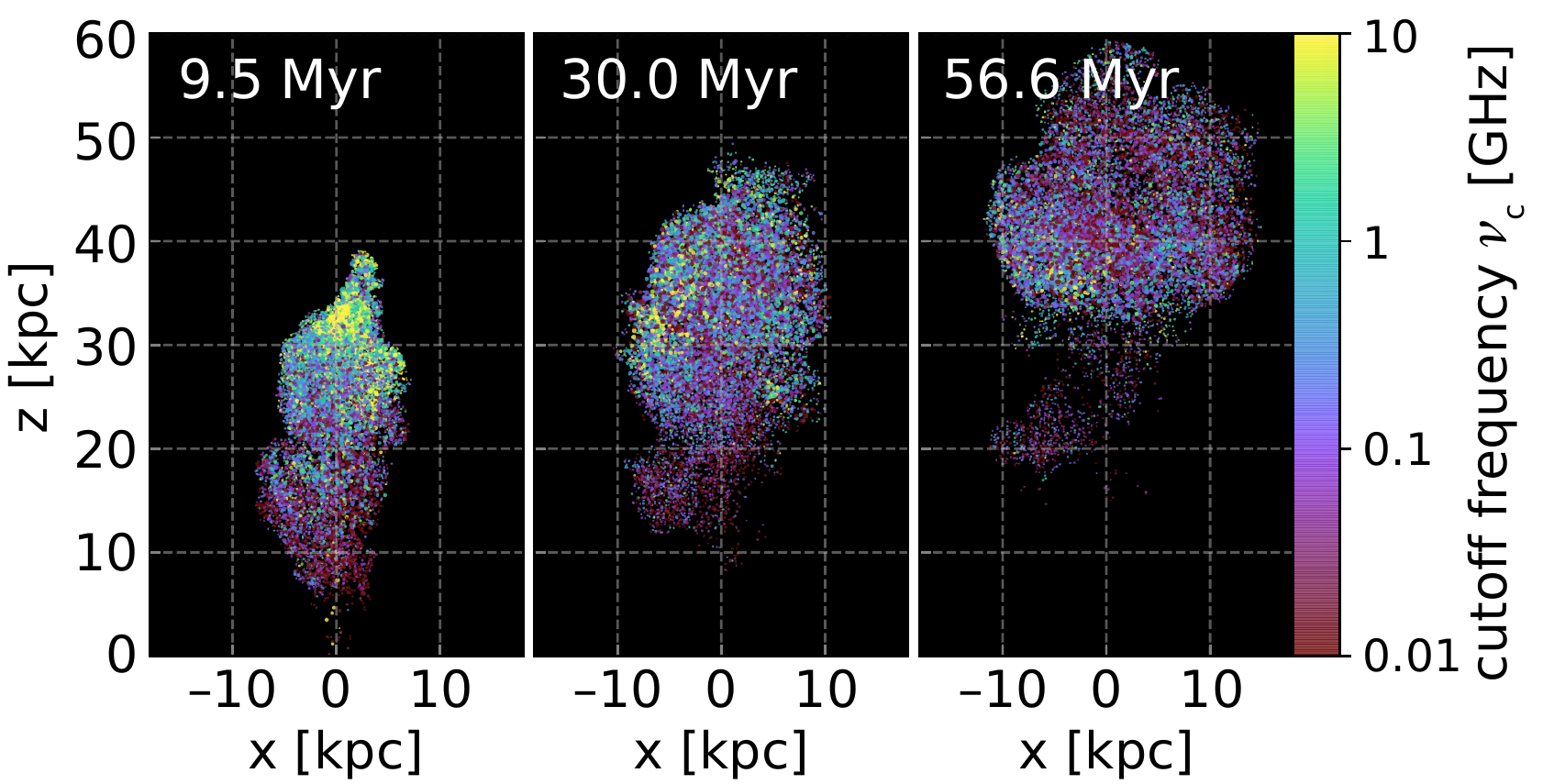}
\caption[Projection of the cutoff frequency of the particles]
	{The cutoff frequency of the \emph{particles} at 3 different times of
	the helical case. The size of the points represents the local gas density--bigger points indicate lower density. While the jet is still active, we can see a clear gradient from the tip to the tail of the lobe. While the jet is off, part of the older plasma that was initially at the tail is lifted by the vortex to the top. This make a \emph{reversed} age gradient at $\sim$ 60 Myr in our case. }
\label{fig:particles_nuc}
\end{figure}

The result of this circulation within the bubble is a reversal of the age
gradient that can be seen at roughly one vortex-turnover time after the jet shuts of, as shown in Fig.~\ref{fig:particles_nuc}. 

This suggests an interesting observational diagnostic, as detecting a reversed age gradient could be used as a dynamical clock. In our simulations, the vortex-turnover time can be estimated using the buoyancy or sound crossing time of the lobe cavity,
\begin{equation}
    t_{\rm vortex}\sim t_{\rm sound}\sim \frac{2R_{\rm lobe}}{c_{\rm s}} \sim 35\,{\rm Myrs}\frac{R_{\rm lobe}}{15\,{\rm kpc}}\frac{900\,{\rm km/s}}{c_{\rm s}}
\end{equation}
which indicates that the vortex should have formed at around the 45Myr mark in our simulations (roughly 35 Myrs after jet turn-off). Because this is a purely dynamical effect, it should be robust against uncertainties in the normalization of the equipartition fraction $\eta$ of the magnetic field (both in the initial conditions and the effects of numerical dissipation), though the optimal spectral region sensitive to measure age gradients will depend on $\eta$.

However, it is important to point out that this effect might still operate in the absence of any buoyancy, at least in part, given the internal circulation driven within the lobe. In particular, \citet{ruszkowski:07} point out that external magnetic fields in the ICM can stabilize buoyant bubbles against disruption, possibly delaying the formation of a buoyant vortex ring. Because both buoyancy and cocoon dynamics act in concert, we expect the reversal of this age gradient to be robust, however, re-simulations will be required that include a magnetized ICM and probe different levels of stratification of the environment to verify this statement.

Confirming the predicted reversal may be observationally challenging. Currently, most observed active FR-II sources exhibit age gradients from the head to the tail since most of the detected sources are higher luminosity objects whose powering jets are still active or have been active recently (and have thus not had time to develop buoyancy-driven vortex rings). Once the jet stops, the radio luminosity quickly drops at least an order of magnitude within a few Myr, which makes detecting the predicted reversal of the age gradient more challenging (since this is a subtle effect, as can be seen in Fig.~\ref{fig:sync_spectralindex}. Although these low-surface brightness radio sources are brighter at low frequencies, the inherently lower resolution at lower frequency might make the detection of the reversing effect more difficult.

To derive a more robust diagnostic, we measure the cumulative flux fraction along the jet
\begin{equation}
    CF_{\nu}(z)=\frac{\int_{-z_{\rm min}}^{z}\int_{y_{\rm min}}^{y_{\rm max}} I_{y,z'} dy dz'}{\int_{-z_{\rm min}}^{z_{\rm max}}\int_{y_{\rm min}}^{y_{\rm max}} I_{y,z} dy dz}
\end{equation}
as a function of distance along the jet at 150MHz and 1.4GHz (such that each is a function between 0 and 1).

\begin{figure*}
\includegraphics[width=\textwidth]{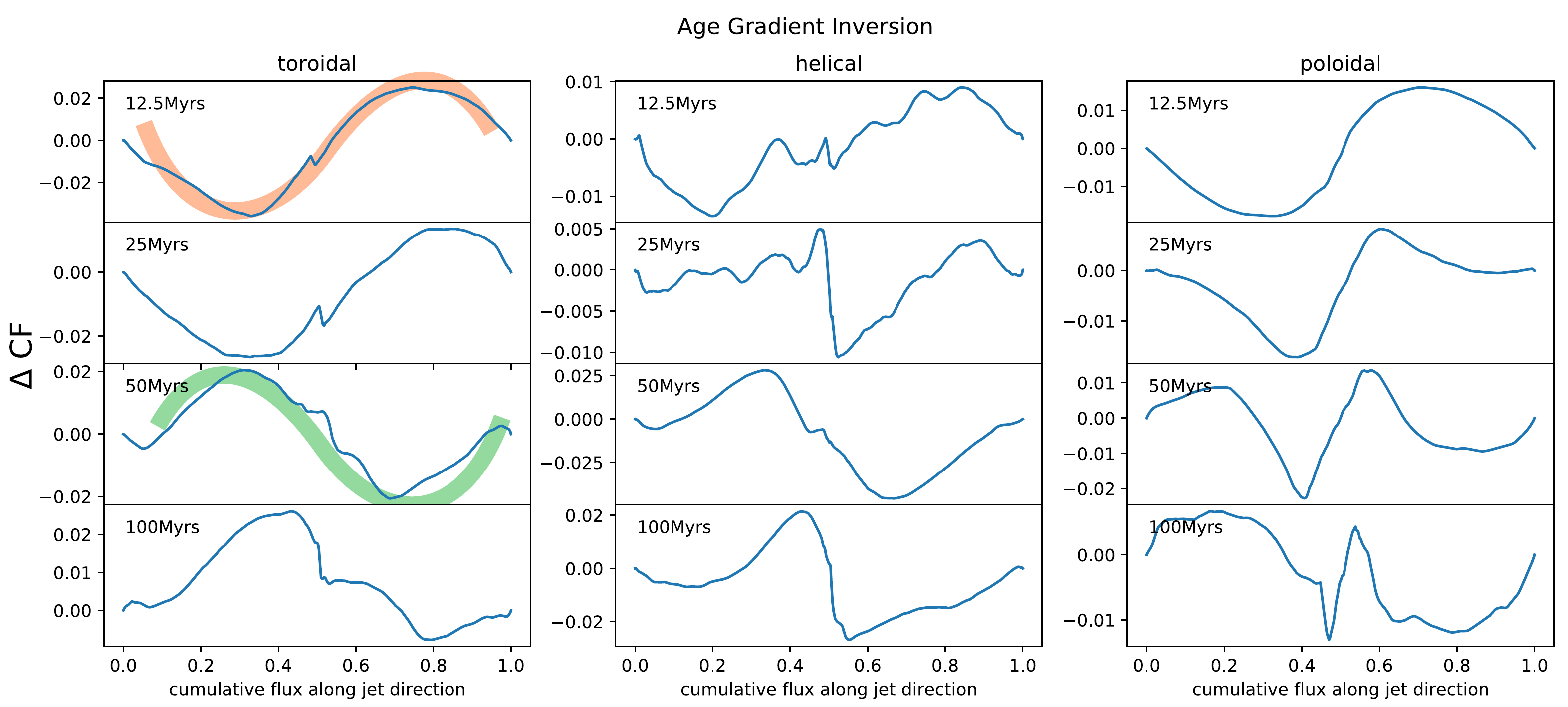}
\caption[Reversal of Age Gradient]
	{Difference in cumulative emission along the jet axis between 150MHz and 1.4GHz, plotted against cumulative flux at 150MHz. All three simulations show a characteristic reversal of the s-shaped curve indicating the reversal of the age gradient in the radio lobes between 25Myrs and 50Myrs. The red and green lines in the left-hand panels show a cartoon of the expected signal of the observed age inversion to highlight the change in curvature. The features at the center of the emission (around $CF\sim 0.5$) in the poloidal case are due to the fact that this case maintains relatively strong emission in the center of the cluster due to the poloidal flux that is preserved even after the jet turns off.}
\label{fig:fractional_offset}
\end{figure*}

Given the observed age gradient, we would expect these $CF_{\nu}(z)$ curves to show a difference in shape depending on whether the emission is more or less centrally concentrated. In Fig.~\ref{fig:fractional_offset}, we then plot the {\em difference} $\Delta CF \equiv CF_{150MHz} - CF_{1.4GHz}$ between these cumulative flux fractions as a function of $CF_{150MHz}$ for the three different simulations and at four different times. At early times (12.5Myrs), all three simulations show an S-shaped curve that indicates an inward age gradient (i.e., more of the high-frequency emission is concentrated at larger distances from the center), thus, the curve shows a valley and then a peak from left to right. However, at later times (50Myrs, 100Myrs), the figures show an overall reversal of this S-shape, indicating the reversal of the age gradient, with younger/high frequency emission concentrated closer to the center. While the detailed shapes of these curves do show some differences, this reversal appears robust and suggests an observational diagnostic that is both simple to derive from multi-frequency data and robust against model assumptions.

Because $CF$ is easy to calculate and presents a twice-integrated measure of intensity, it is relatively robust {{and substantially less noisy than, for example,}} spatial averages of the spectral index as a function of $z$. We propose using this diagnostic as a measure of spatially resolved spectral aging.

{{More work beyond the scope of this publication will be required to establish the presence of this reversal age gradient and to calibrate $CF$ as a true universal diagnostic. Still, the promise to quantify the time since injection of particles by active jets ceased (i.e., the time since the jets shut off) would be useful in, for example, determining the fraction of time radio sources inflated in galaxy clusters are actively powered vs.~passively evolving, this providing a new constraint on the duty cycle of the jet. Conversely, since the synchrotron cooling age depends on $\eta$, establishing the actual dynamical age of the plasma (one turnover time since the jet turned off) could be used to constrain $\eta$.}}

%\begin{figure}
%\includegraphics[width=\columnwidth]{fractional_offset_z_14%00_100.pdf}
%\caption[Fractional offset of the centers between 1400MHz and %100MHz]
%	{Fractional offset of the weighted centers between 1400MHz and %100MHz. The emission weighted distance from the central AGN are %calculated from two frequencies and the fractional difference is %shown here. The offset is positive during time when the jet is %active. It remains positive for roughly a turnover time and
%	disappears after 60 Myr.}
%\label{fig:fractional_offset}
%\end{figure}

%\begin{minipage}{\textwidth}
%\end{minipage}

\subsection{Caveats and Limitations of Our Simulation}

While numerical simulations of jet propagation and radio galaxy evolution are proliferating in the literature, each work must necessarily make simplifying assumptions and suffers from unavoidable numerical limitations. It is therefore important to discuss the limitations and caveats of our work in this context.

\subsubsection{Strength of the magnetic fields in the jet and lobes, numerical dissipation}
\label{sec:strength_of_Bfields}

In our simulations, we do not set the strength of the magnetic fields explicitly. Instead, we choose the plasma beta $\beta_p$ to be of order unity inside the jet nozzle, which implies that the magnetic pressure is comparable to the thermal pressure of the gas and is dynamically important {\em within the jet}. The field strength is then determined by our choices of jet power, internal Mach number, and nozzle cross-section. We note the resulting magnetic field strength at the jet base ($\sim$ kpc from the black hole) is about 170 $\mu$G. While it is currently impossible to precisely infer the strength of the magnetic fields in the jets from observations, \cite{osullivan:09} find the magnetic fields in AGN cores are in the range of $100 \sim 300$ $\mu$G at 1 pc and scales like $r^{-1}$. Our setting of magnetic fields is consistent with their findings.

However, we inject our jets ballistically, i.e., assuming most of the magnetic and thermal energy has been converted to kinetic energy on scales smaller than the boundaries of our injection nozzle roughly about 100 pc from the black hole. This is one possible assumption about the nature of jet launching and propagation. For example, \cite{sikora:05} argue that, from observational data, the conversion of Poynting flux to kinetic energy happens mostly within $10^3$ gravitational radii at parsec scale, which is consistent with our assumptions. However, other plausible models exist which maintain mostly poynting-flux dominated jets even on kpc scales \citep[e.g.,][]{colgate:15}, which are clearly in conflict with the assumptions underlying our simulations.

The relatively small fraction of thermal and magnetic energy in the jets accounts for the relatively minor differences between different magnetic topologies in the overall shapes of the sources, since the kinetic power is approximately two orders of magnitude larger than the thermal and electro-magnetic power. This also implies that, while the jets are injected with a plasma beta of order unity, the shock-heating of jet fluid substantially increases the plasma beta within the lobes, i.e., the field strength in the radio lobes is well below equipartition.

In our current setup, the surrounding medium is non-magnetized. All the magnetic flux comes from the injection of the jet. Although this setup is obviously unphysical, we use our simulations to create a baseline for future studies in which more complicated factors might affect the dynamics of the jets and the bubbles. Some research hints that the evolution of bubbles might be affected more by the surrounding magnetic field than by the field inside the bubbles \citep{ruszkowski:07,dursi:08}. We will leave the investigation of jets interacting with magnetized surroundings to future work.

Finally, it is important to address the question of numerical diffusivity, which is an unavoidable consequence of any Eulerian scheme to integrate the MHD equations (note once again that FLASH preserves the solenoidal condition to machine accuracy). The effect of this can be seen in Fig.~\ref{fig:mag_energy}, which shows that the magnetic energy of the simulations is substantially reduced at late times. 

Some reduction in the total magnetic energy is expected because of the PdV work done by the magnetic field (as part of its adiabatic evolution). The reduction in the poloidal case is of order 50\%, which is consistent with the expectation of simple adiabatic behavior, given that the radio lobes expand roughly by a factor of 2 in radius over this timescale. 

However, the reduction by up to 80\% is substantially in excess of that expectation in the helical and toroidal cases. We attribute this reduction to the numerical reconnection of magnetic energy on small simulation scales in the radio lobes, which is more prevalent in the toroidal and helical cases, given the much shorter field-reversal scales. Comparing to the poloidal case, we conclude that of order 50\% of the field energy in the helical and toroidal simulations is lost to numerical reconnection, beyond the level of turbulent reconnection we would naturally expect to occur \citep[e.g.][]{lazarian:20}. Reducing the magnitude of this effect would be desirable, but would require a substantially higher resolution, beyond the scope of resources available at the time of the simulations. 

As such, aspects of our simulations subject to the total magnitude of the magnetic field strength (such as the absolute value of the spectral break/cutoff frequency, see below, and the absolute value of the synchrotron intensity) should be considered with these caveats in mind, while qualitative results, such as the overall spatial distribution of age gradients and the differences in emission morphology between simulations will be more robust against uncertainties introduced by numerical dissipation. Because FLASH is an energy conserving code, any effects of numerical dissipation simply increase the internal energy of the gas.

\subsubsection{On Spectral Aging Models}

Conventionally, spectral aging models assume a constant magnetic field, which is usually derived from minimum energy arguments, to calculate the age of the radio source. However, as seen in our simulations, the magnetic fields that a particular plasma ensemble experiences along its trajectory are time variable. In the hotspots where the fresh plasma begins to cool, the magnetic fields are strongest due to compression at the terminal shocks. When the plasma leaves the hotspots, the magnetic field strength significantly drops because of the adiabatic expansion of the plasma as well as reconnection of the highly tangled field. We note that this latter effect is enhanced in any numerical simulation because of limited resolution and numerical diffusivity.

As a consequence, the cooling rates in the lobes are much lower compared to the cooling rates in the hotspots. The evolution of the spectral index depends on the strength of the magnetic field, due to the reduction in overall field strength from lobe expansion and numerical reconnection.

The Lagrangian particles begin cooling after they leave the jet. Our current model does not capture any re-acceleration outside the hotspots. A more complex framework \citep[e.g.][]{vaidya:18} would be necessary to study various acceleration mechanisms and the resulting synchrotron emission.

\section{Conclusions}
\label{sec:conclusions}

In this paper we present a series of MHD simulations of AGN jets with different injecting field topology. We set up pure toroidal, pure poloidal, and helical magnetic fields inside the jets (see Fig.~\ref{fig:mag_drawing}). The jets are then evolved according to the ideal MHD formalism. To this end, we implement an implicit scheme to integrate the particle transport equation that allows spectral age modeling of radio plasmas.

Our simulations reproduce the well-known result that non-relativistic, heavy jets cannot reproduce realistic radio lobe shapes unless a random jitter---i.e., a dentist drill effect--- is imposed on the orientation of the jet axis.

We find that the presence of toroidal magnetic fields reduces the lateral expansion of the jet and enables faster propagation of the jet heads. On the other hand, the presence of poloidal fields inhibits the development of kink instability. Thus, we find that purely toroidal jets do not propagate faster than helical jets, despite the stronger effects of self-collimation at equal field strength. Instead, their propagation speed is comparable to the helical case, as the kink-instability effectively spreads the momentum flux of the jet over a larger solid angle than might be expected from a fully self-collimated jet.

We find that, due to the more efficient collimation, toroidal and helical jets in our simulations generate more elongated cocoons with more pronounced hourglass morphology and that, as a result, cavities inflated by these jets are further away from the central engine compared to those inflated by the poloidal and hydrodynamic jets in our simulations.

While the initial injection of magnetic energy into the lobes reflects the topology of the magnetic field in the jet, turbulence within the lobes quickly re-arranges the field to rough equipartition between the different field orientations.

We find that, after the jets turn off, the radio lobes in our simulations develop the well-known rising vortex ring structure of buoyant bubbles in stratified atmospheres in all cases. When viewed from lines of sight close to the jet axis, the toroidal structure of the emission is readily observable.

We show that the large scale rotation of the vortex rings leads to the a reversal of the spectral index gradient after the jet shuts off, due to the reciprocal motions of the young and old plasma. This suggests that the observation of spectral age gradients can be used as a dynamical clock. While this effect is likely most pronounced in FR-II radio galaxies (most similar to the radio sources produced by our simulations), some FR-I radio galaxies should show similar reversals. We propose a simple, robust diagnostic to search for such age gradient signatures.

\section{Data Availability}
\label{sec:data}

The numerical data and data products presented in this paper are hosted on storage maintained by the Department of Astronomy at the University of Wisconsin-Madison. Data and products will be shared on reasonable request to the corresponding author, Sebastian Heinz, at sheinz@wisc.edu.

\end{document}